\documentclass[journal,10pt]{IEEEtran}
\usepackage{amsmath,amssymb}
\usepackage{graphicx}
\usepackage{booktabs}
\usepackage{algorithmic}
\usepackage{cite}
\usepackage[hyphens]{url}
\usepackage{microtype}
\graphicspath{{figures/}{upload/}}

\newif\ifextendedfigures
\extendedfiguresfalse

\DeclareMathOperator{\symlog}{symlog}
\DeclareMathOperator{\symexp}{symexp}
\DeclareMathOperator{\Var}{Var}
\newcommand{\pospart}[1]{\left[#1\right]_+}

\newcounter{rtwmalg}

\begin{document}
\raggedbottom
\title{RTWM: Real-Trace World Models for Multi-UAV Edge Computing Networks}
\author{Shengjie Zhong,~\IEEEmembership{Graduate Student Member,~IEEE,}
        Zhongliang Zhao,
        Jingxuan Chen,
        Xianbin Cao,
        Kaiting Zhang,
        Dapeng O. Wu,~\IEEEmembership{Fellow,~IEEE},
        Tony Q. S. Quek,~\IEEEmembership{Fellow,~IEEE}
\thanks{This work was supported by the Smart Grid-National Science and Technology Major Project (2026ZD0811400). (Corresponding author: Zhongliang Zhao.)}
\thanks{S. Zhong is with the School of Electronic and Information Engineering, Beihang University, Beijing 100191, China (e-mail: zhongshengjie@buaa.edu.cn).}
\thanks{Z. Zhao is with the School of Electronic and Information Engineering, Beihang University, Beijing, with the Department of Strategic and Advanced Interdisciplinary Research, Peng Cheng Laboratory, with the Hangzhou International Innovation Institute, Beihang University, Hangzhou, with the MIIT Key Lab. of Aerospace Mobile Communication, Beijing, and also with the State Key Lab. of CNS/ATM, Beijing (e-mail: zhaozl@buaa.edu.cn).}
\thanks{J. Chen is with the Department of Computer Science, City University of Hong Kong, Hong Kong, SAR, China, and also with the Hong Kong Generative AI Research and Development Center, Hong Kong, SAR, China (e-mail: chenjingxuan@buaa.edu.cn).}
\thanks{X. Cao is with the School of Electronic and Information Engineering, Beihang University, Beijing, and also with the MIIT Key Lab. of Aerospace Mobile Communication, Beijing.}
\thanks{K. Zhang is with the School of Electronic and Information Engineering, Beihang University, Beijing 100191, China (e-mail: zhangkaiting@buaa.edu.cn).}
\thanks{D. O. Wu is with the Department of Computer Science, The City University of Hong Kong, Hong Kong, SAR, China (e-mail: dpwu@ieee.org).}
\thanks{T. Q. S. Quek is with the Information Systems Technology and Design Pillar, Singapore University of Technology and Design, Singapore 487372 (e-mail: tonyquek@sutd.edu.sg).}
}

\markboth{IEEE trans, Vol.~XX, No.~XX, XXX~20XX}%
{Zhong et al.: RTWM for Multi-UAV Edge Computing Networks}
\IEEEspecialpapernotice{This work has been submitted to the IEEE for possible publication. Copyright may be transferred without notice, after which this version may no longer be accessible.}
\maketitle

\begin{abstract}
Multi-UAV mobile edge computing couples motion, bandwidth, CPU-frequency, and offloading decisions under stochastic arrivals and per-task deadlines. Because demand and user mobility evolve regardless of fleet actions, online control must separate what cannot be influenced from what can and replan in real time. Existing world-model controllers learn a single action-conditioned transition, spending capacity on exogenous variation that cannot change action rankings, and are usually validated on synthetic demand. This paper presents the real-trace world model (RTWM), a planning framework built around a trace-calibrated factorized world model for deadline-aware UAV edge computing, calibrated on three city-scale traces replayed at 10-minute resolution. Action-free ensembles forecast the demand field and representative-user dynamics; an action-conditioned component propagates service outcomes, queue backlogs, motion, and batteries for all candidate sequences over one shared exogenous rollout; and a reset-mean cross-entropy planner with ensemble-disagreement penalty and terminal value selects an action at every 600 s slot under fixed candidate, horizon, and CEM-iteration budgets. On regular days it has the highest observed mean deadline satisfaction in two of three cities and remains close elsewhere.
\end{abstract}


\section{Introduction}
\IEEEPARstart{M}{ulti}-unmanned-aerial-vehicle (UAV) mobile edge computing (MEC) is emerging as a supporting technology for 6G space--air--ground integrated networks and large-scale Internet-of-Things (IoT) service~\cite{cao2024computational,zhou2025cooperative}. By placing computing and communication payloads on rotary-wing UAVs, operators can extend edge capacity on demand to infrastructure-poor areas and post-disaster scenes~\cite{cao2024computational}. Serving real workloads with such a fleet, however, requires jointly deciding where the UAVs fly, how the fleet shares its bandwidth and CPU frequency, and which tasks are offloaded, while tasks arrive stochastically and expire once their deadlines pass~\cite{bai2024delay,hao2024joint}. Rotor propulsion and battery cycling couple every one of these choices to energy~\cite{zeng2019energy}, and dynamic voltage and frequency scaling (DVFS) ties computation power to the serving decisions themselves~\cite{wang2016mobileedge}. The resulting control problem is high-dimensional and strongly coupled, subject to random arrivals and hard timing constraints, and must be resolved online. \textbf{How can a UAV-MEC controller sustain a high deadline satisfaction ratio (DSR) while containing fleet energy and per-slot decision cost, under stochastic arrivals and tightly coupled motion, communication, and computation decisions?}

World models offer a promising route to this goal. Instead of learning a reactive policy, they learn a predictive latent dynamics and plan inside it~\cite{ha2018world,hafner2019learning}. They have recently matched or surpassed model-free reinforcement learning across manipulation, locomotion, navigation, and driving benchmarks~\cite{hafner2025mastering,hansen2024tdmpc2}. Their appeal for UAV-MEC is sample efficiency: a controller can rehearse candidate action sequences in imagination rather than collecting years of interaction with a costly simulator. The standard recipe, however, trains a single transition network conditioned on the agent's action, implicitly assuming that the environment's evolution is action-driven. That assumption fails in a structurally important way in UAV-MEC. The demand field and the positions, activity, and demand shares of end users evolve exogenously---no UAV action changes them---whereas service rates, queues, motion, and batteries respond directly to control. A monolithic world model therefore spends predictive capacity on exogenous signals that cannot alter the ranking of candidate actions, and it entangles them with the action-conditioned dynamics that actually determine control quality. The RTWM framework removes this mismatch by construction: action-free exogenous ensembles forecast the demand field and representative-user states, while explicit, differentiable physics propagates service, queues, motion, and batteries for each candidate action sequence.

We formalize deadline-aware multi-UAV edge computing as a finite-horizon Markov decision process over a fleet of $K$ rotary-wing UAVs serving $M$ representative end users (EUs), calibrated on three city-scale demand traces: Milan Internet-CDR counts~\cite{barlacchi2015multisource}, Shanghai session initiations~\cite{mexwell2023telecom,li2022profit,guo2020user,wang2021delay}, and YJMob distinct-user presence~\cite{yabe2024yjmob100k}, all replayed at 10-minute resolution. Demand and mobility fields are only partially predictable from history~\cite{gong2025sttf,he2025mamba}, so rather than assuming a parametric generative model of user behavior, we calibrate the exogenous activity fields and EU counts directly from each trace and hold them fixed across candidate actions during planning. The simulation stack models probabilistic line-of-sight air-to-ground channels~\cite{alhourani2014optimal}, DVFS-aware CPU energy~\cite{wang2016mobileedge}, rotary-wing propulsion with return-to-station charging~\cite{zeng2019energy}, task-level FIFO queues, and deadline expiry. Every method in this paper is evaluated in a trace-replay closed loop in which the exogenous fields are shared across candidate action sequences while service and physical states are propagated per candidate (Sections~III and~V).

Building on this formulation, the paper makes three contributions. It develops a complete trace-replay simulation stack for deadline-aware multi-UAV MEC that replays exogenous demand and user fields while exposing motion, bandwidth, CPU-frequency, and offloading controls under realistic energy, channel, and deadline constraints. It introduces RTWM, a control framework that couples a trace-calibrated factorized world model---action-free exogenous ensembles feeding action-conditioned service, queue, motion, and battery propagation, trained with multi-step losses---with reset-mean cross-entropy planning, an ensemble-disagreement penalty, and a terminal critic value under receding-horizon replanning. It further reports a systematic paired comparison against heuristic, model-free, and model-based baselines across three datasets and operating regimes spanning the demand multiplier, the EU count, and the fleet size, stating the boundaries as plainly as the gains: on YJMob regular days RTWM's observed mean trails only the INS-WOA heuristic among the non-learned references and is on par with the TD-MPC-style comparator, the balanced 32-UAV configuration stays above both heuristics, and its Milan open-loop forecasts are weaker than persistence. We also note that evaluation days were reused across twenty development variants, so the exploratory paired intervals reported throughout are conditional on the data and method selections described herein. The remainder of the paper is organized as follows: Section~II reviews related work, Section~III formalizes the system, Section~IV develops RTWM, Section~V presents experimental results, and Section~VI concludes.

\section{Related Work}
\subsection{From Analytical Optimization to Metaheuristics in UAV-MEC}
Early work in UAV-MEC formulated control as mathematical programming, and the line remains active: joint optimization of trajectory, offloading, caching, and migration~\cite{zhao2025joint}, trajectory design and resource allocation for integrated sensing, communication, and edge computing~\cite{peng2025trajectory}, and two-timescale joint control of mobility and computation~\cite{sun2025tjcct}. Related formulations also jointly optimize data acquisition and trajectory planning for UAV-assisted wireless-powered IoT systems~\cite{ning2025joint}. Offloading and trajectory co-design under dynamic users~\cite{zhang2024task} and multi-objective delay--energy optimization~\cite{sun2024multiobjective} continue the formulation line. Joint communication--computation optimization for latency and energy~\cite{pervez2024energy,guo2024multiuav} completes the formulation layer. Deadline awareness entered through delay-aware cooperative offloading across edge--cloud clusters~\cite{bai2024delay} and priority-aware admission and scheduling~\cite{hao2024joint}. Reliability-aware formulations add probabilistic guarantees~\cite{hao2025reliability}. A complementary line asks where edge capacity should sit, optimizing server placement and user allocation for profit, coverage, or delay~\cite{li2022profit,guo2020user,wang2021delay,wu2025fairness}. Learning-based DVFS ties these models to hardware energy knobs~\cite{zhang2024dvfo,wang2016mobileedge}. Because closed-form solutions exist only under simplifications, metaheuristics---the whale optimization algorithm and its non-dominated-sorting variants, particle swarm optimization, and genetic algorithms---remain the default search machinery. Recent instances include joint trajectory--resource--access designs for the low-altitude economy~\cite{li2026joint} and hierarchical UAV swarms in 6G aerial access networks~\cite{mirjalili2016whale,jia2026dynamic}. Dynamic computation-rate maximization~\cite{chen2025dynamic} follows the same pattern. These methods rely on accurate analytical channel and energy models and are solved offline or per slot; under stochastic arrivals, deadline expiry, and a joint action space spanning motion, bandwidth, CPU frequency, and offloading, their computational structure scales poorly to real-time fleet control~\cite{cao2024computational}.

\subsection{Deep Reinforcement Learning and Emerging Backbones}
Deep reinforcement learning (DRL) sidesteps tractable models by learning control policies directly. Generative diffusion models have also been surveyed as optimization machinery for network problems~\cite{du2024enhancing}. Single-agent algorithms have been applied to joint UAV positioning and offloading~\cite{schulman2017proximal,wang2025joint}. Maximum-entropy variants such as SAC improve exploration stability for offloading under dynamic channels~\cite{haarnoja2018soft,deng2024entropy}. Multi-agent extensions coordinate fleets under centralized training with decentralized execution: cooperative MADDPG with learned communication~\cite{lowe2017multiagent,kim2024cooperative} and decentralized multi-agent designs for multi-UAV MEC networks~\cite{hwang2025multiagent}. Federated training improves trajectory control and fairness~\cite{tarekegn2025trajectory}, and attention-augmented variants handle rendezvous in dynamic obstacle fields~\cite{xing2026multiuav}. Graph representation learning structures these agents further, from graph-attention trajectory and resource assignment~\cite{feng2024graphattention} to deep graph reinforcement learning for secure UAV communications~\cite{tang2025deep} and GNN--MARL foundations~\cite{liu2024graph}. In parallel, sequence architectures have entered networking research. Spatiotemporal Transformers improve multi-task mobile-network prediction~\cite{gong2025sttf}, and state-space models such as Mamba are being explored for long-horizon traffic forecasting~\cite{he2025mamba}. Large language models have also been explored for autonomous edge intelligence~\cite{shen2024large}. Digital twins are being proposed as external planners and oracles for edge systems~\cite{zhou2025cooperative}. These are parallel explorations rather than components of this paper. What DRL methods share is the absence of an explicit predictive model: they learn reactive mappings without modeling queues, batteries, or exogenous demand, which costs sample efficiency and offers no mechanism to share exogenous forecasts across candidate actions, the gap that world models are designed to fill.

\subsection{World Models and Model-Based Planning}
World models invert the DRL strategy: learn the dynamics, then plan inside them. The paradigm originates from recurrent world models trained by policy evolution~\cite{ha2018world}, moves to latent dynamics learned from pixels with planning via the cross-entropy method in PlaNet~\cite{hafner2019learning}, and matures in the Dreamer line of agents, whose latest generation masters over 150 diverse control tasks with a single fixed configuration~\cite{hafner2025mastering}. Temporal-difference learning merged with MPC in TD-MPC and its successor, replacing reconstruction losses with value-consistent latent prediction and short-horizon online planning~\cite{hansen2022temporal,hansen2024tdmpc2}. Ensemble disagreement is an increasingly common signal for deciding where imagined rollouts can be trusted~\cite{frauenknecht2024trust}, and sampling-based MPC such as MPPI supplies robust trajectory optimizers for real aerial vehicles~\cite{williams2017information,andrejev2025pimppi}. Beyond simulated benchmarks, world models are being transferred to navigation and driving~\cite{guan2024world} and, very recently, to wireless problems, with Dreamer-style models used for long-term age-of-information minimization in vehicular networks~\cite{wang2025world}. What these works share is the assumption that the learned transition is action-conditioned. That assumption is structurally mismatched with UAV-MEC, where demand and user mobility evolve independently of control. RTWM is positioned at this gap: the framework factorizes its world model into action-free exogenous ensembles and action-conditioned service, queue, motion, and battery propagation, calibrates the exogenous state on replayed city-scale traces, and wraps the result in reset-mean CEM replanning with disagreement penalty and terminal value---a combination designed for stochastic, deadline-driven edge fleets rather than for action-driven simulators.

\section{System Model and Problem Formulation}
\label{sec:system_model}

\subsection{UAV-MEC Network and Control}
\begin{figure}[t]
\centering
\includegraphics[width=0.95\columnwidth]{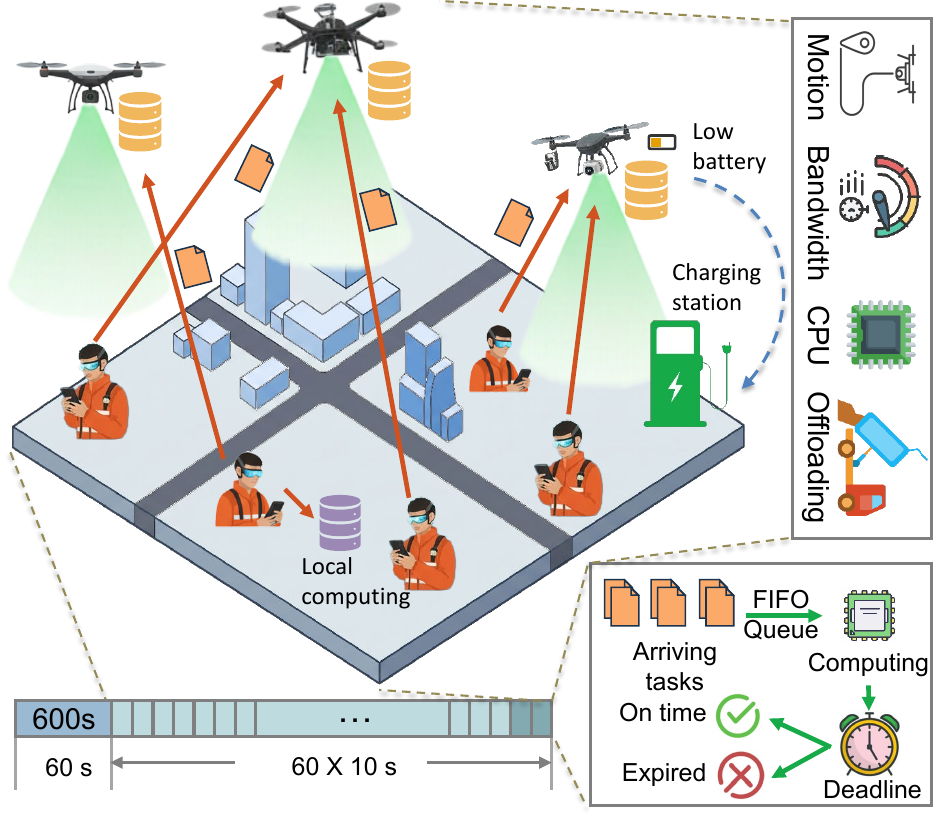}
\caption{UAV--MEC simulator. Replayed demand and representative end-user fields drive task queues over 10-s subslots. Every 600~s, the controller sets motion, bandwidth, CPU frequencies, and offloading probabilities.}
\label{fig:scenario}
\end{figure}

A fleet $\mathcal K=\{1,\ldots,K\}$ of rotary-wing UAVs serves $\mathcal A=[0,L]^2$, with $L=G\ell_{\mathrm{cell}}$ for a $G\times G$ grid of cell width $\ell_{\mathrm{cell}}$. Here $K=16$ unless specified otherwise. UAV $u$ has horizontal position $\mathbf p_{u,t}$ and altitude $h_{\mathrm{UAV}}=100$ m. Each decision slot lasts $\Delta=600$ s and contains $N_s=60$ service subslots of duration $\delta=10$ s. Fig.~\ref{fig:scenario} summarizes the simulator.

The simulator maintains $M=32$ representative end users (EUs) by default. At slot $t$, EU $m$ has replayed position $\mathbf x_{m,t}$, binary activity $\zeta_{m,t}$, and demand share $w_{m,t}\geq0$. With $N_t=\sum_m\zeta_{m,t}$, the weights satisfy
\begin{equation}
 \sum_{m=1}^{M} w_{m,t}=\mathbb I\{N_t>0\},\qquad
 \zeta_{m,t}=0\Rightarrow w_{m,t}=0.
 \label{eq:eu_weights}
\end{equation}
These synthetic EUs represent shares of aggregate activity. The environment replays their positions, activities, and weights independently of UAV actions.

The joint action $\mathbf a_t\in[-1,1]^{4K+M}$ contains UAV controls $(a^x_{u,t},a^y_{u,t},a^b_{u,t},a^f_{u,t})$ and EU offloading controls $a^o_{m,t}$, with $o_{m,t}=(a^o_{m,t}+1)/2$. For an airborne UAV, let $\mathbf v^{\mathrm{cmd}}_{u,t}=(a^x_{u,t},a^y_{u,t})^{\mathsf T}$. Then
\begin{align}
 \Delta\mathbf p^{\mathrm{cmd}}_{u,t}
 &=\frac{D_{\max}\mathbf v^{\mathrm{cmd}}_{u,t}}
 {\max\{1,\|\mathbf v^{\mathrm{cmd}}_{u,t}\|_2\}},\nonumber\\
 \mathbf p_{u,t+1}&=\operatorname{clip}_{\mathcal A}
 (\mathbf p_{u,t}+\Delta\mathbf p^{\mathrm{cmd}}_{u,t}),
 \label{eq:uav_motion}
\end{align}
where $D_{\max}=1500$ m. Returning UAVs move autonomously toward a station. Charging UAVs remain stationary and offline. The simulator applies motion at the slot boundary and serves from the post-motion position throughout the slot.

Let $\mathcal F_t$ contain airborne and returning UAVs, and set $\chi_{u,t}=\mathbb I\{u\in\mathcal F_t\}$, where $\mathbb I$ is the indicator function. Fleet bandwidth and per-UAV CPU frequency are
\begin{align}
 B_{u,t}&=B_{\mathrm{tot}}\frac{\chi_{u,t}\exp(a^b_{u,t})}
 {\sum_{v\in\mathcal K}\chi_{v,t}\exp(a^b_{v,t})},\nonumber\\
 f_{u,t}&=\chi_{u,t}\frac{a^f_{u,t}+1}{2}F_{\max}.
 \label{eq:resource_allocation}
\end{align}
Equation~\eqref{eq:resource_allocation} applies for $\mathcal F_t\ne\varnothing$. All allocations are zero when $\mathcal F_t=\varnothing$. When UAVs are online, the fleet divides $B_{\mathrm{tot}}=20$ MHz into orthogonal shares. Each online UAV has an independent processor with $F_{\max}=10$ GHz. The model assumes ideal time or frequency sharing within each UAV and omits inter-UAV interference.

\subsection{Links, Association, and Service Rates}
With horizontal distance $d^{\mathrm{hor}}_{m,u,t}=\|\mathbf x_{m,t}-\mathbf p_{u,t+1}\|_2$ and slant range $d^{\mathrm{slant}}_{m,u,t}=\sqrt{(d^{\mathrm{hor}}_{m,u,t})^2+h_{\mathrm{UAV}}^2}$, the implemented probabilistic-LoS approximation~\cite{alhourani2014optimal} uses
\begin{equation}
 \theta_{m,u,t}=\frac{180}{\pi}\operatorname{atan2}
 \bigl(h_{\mathrm{UAV}},\max\{d^{\mathrm{slant}}_{m,u,t},1~\mathrm m\}\bigr).
 \label{eq:implemented_angle}
\end{equation}
Both the simulator and differentiable model use this slant-range approximation, which gives $45^\circ$ directly below a UAV. The LoS probability is
\begin{equation}
 P^{\mathrm{LoS}}_{m,u,t}=\frac{1}{1+A_{\mathrm{LoS}}
 \exp[-B_{\mathrm{LoS}}(\theta_{m,u,t}-A_{\mathrm{LoS}})]},
 \label{eq:los_probability}
\end{equation}
where $(A_{\mathrm{LoS}},B_{\mathrm{LoS}})=(9.61,0.16)$. Average path loss and Shannon spectral efficiency are
\begin{align}
 \operatorname{PL}_{m,u,t}={}&20\log_{10}\!\left(
 \frac{4\pi f_cd^{\mathrm{slant}}_{m,u,t}}{c_0}\right)\nonumber\\
 &+P^{\mathrm{LoS}}_{m,u,t}\eta_{\mathrm{LoS}}
 +(1-P^{\mathrm{LoS}}_{m,u,t})\eta_{\mathrm{NLOS}},\nonumber\\
 q_{m,u,t}={}&\log_2\!\left(1+\frac{P_{\mathrm{tx}}
 10^{-\operatorname{PL}_{m,u,t}/10}}{N_0B_{\mathrm{tot}}}\right),
 \label{eq:link_quality}
\end{align}
with $f_c=2$ GHz, $c_0=3\times10^8$ m/s, $P_{\mathrm{tx}}=0.5$ W, $N_0=10^{-20.4}$ W/Hz, and $(\eta_{\mathrm{LoS}},\eta_{\mathrm{NLOS}})=(1,20)$ dB. The noise model omits receiver noise figure and implementation margin.

Each active EU associates with its nearest two non-charging UAVs, or all available UAVs if fewer than two remain; denote this set by $\mathcal U_{m,t}$. Link quality and UAV load determine
\begin{align}
 \phi_{m,u,t}&=\frac{q_{m,u,t}}{\sum_{v\in\mathcal U_{m,t}}q_{m,v,t}},\qquad
 L_{u,t}=\!\!\sum_{\substack{m:\zeta_{m,t}=1\\u\in\mathcal U_{m,t}}}
 \phi_{m,u,t}w_{m,t},\nonumber\\
 s_{m,u,t}&=\frac{\phi_{m,u,t}w_{m,t}}{L_{u,t}},\nonumber\\
 r_{m,t}&=\!\!\sum_{u\in\mathcal U_{m,t}}\!\!q_{m,u,t}B_{u,t}s_{m,u,t},\nonumber\\
 f^{\mathrm{edge}}_{m,t}&=\!\!\sum_{u\in\mathcal U_{m,t}}\!\!f_{u,t}s_{m,u,t}.
 \label{eq:association_and_rates}
\end{align}
Load normalization prevents multiple EUs from claiming the same capacity. If $\mathcal U_{m,t}=\varnothing$, then $r_{m,t}=f^{\mathrm{edge}}_{m,t}=0$; a zero link-quality sum gives $\phi_{m,u,t}=0$, and $L_{u,t}=0$ gives $s_{m,u,t}=0$.

The aggregate rates serve one offload FIFO per EU. The model permits a head task to use both UAVs under ideal divisible service and abstracts away partitioning, coordination, recombination, and result downlink.

\begin{figure*}[t]
\centering
\includegraphics[width=\textwidth]{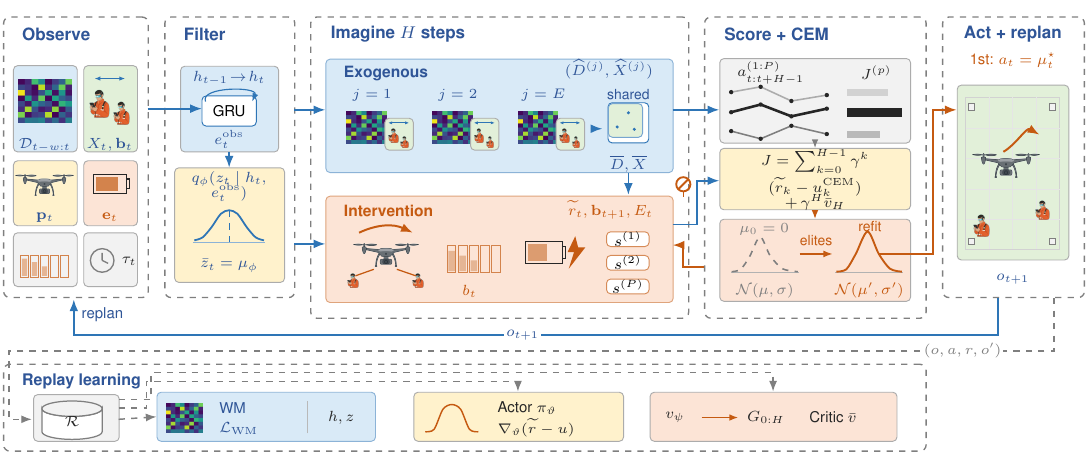}
\caption{RTWM filtering, imagination, and control. Candidates share action-free field predictions and maintain separate service, queue, motion, and battery states. Replay supplies training data for the world model, actor, and critic.}
\label{fig:framework}
\end{figure*}

\subsection{Tasks, Deadlines, and Queues}
Let $D_t(g)$ be the demand field scaled by the training-set 99th percentile of aggregate demand. Its sum is $S_t=\sum_g D_t(g)$. With demand multiplier $\xi$ (default $1$), the arrival rate and subslot task count are
\begin{equation}
 \lambda_{m,t}=\lambda^\star\xi S_t w_{m,t},\qquad
 n_{m,t,s}\sim\operatorname{Poisson}(\lambda_{m,t}\delta).
 \label{eq:task_arrivals}
\end{equation}
Training-only calibration sets $\lambda^\star$ for a peak utilization of 0.8, accounting for serial service and charging duty. A task $i$ arriving at $T_i^{\mathrm{arr}}$ draws payload $D_i\sim\mathcal U[0.5,2.0]$ Mbit and compute density $c_i\sim\mathcal U[500,1500]$ cycles/bit, giving $C_i=D_ic_i$ with $D_i$ expressed in bits. Its deadline is $T_i^{\mathrm{due}}=T_i^{\mathrm{arr}}+\tau_i$, where $\tau_i=\delta\lceil\widetilde\tau_i/\delta\rceil$ and $\widetilde\tau_i\sim\mathcal U[10,50]$ s. These task attributes are synthetic.

Each EU maintains local and offload FIFO queues. With a usable link, a task enters the offload queue with probability $o_{m,t}$. Tasks assigned to local service and tasks without a usable link execute at $f_{\mathrm{loc}}=1$ GHz. Offloaded tasks transmit before computing. For the head task, let $x_i$ and $y_i$ be remaining bits and cycles (initially $D_i$ and $C_i$), and let $\delta_i\leq\delta$ be the remaining time in the subslot. When rates are positive,
\begin{align}
 s_i^{\mathrm{tx}}&=\min\{\delta_i,x_i/r_{m,t}\},\nonumber\\
 s_i^{\mathrm{cp}}&=\min\{\delta_i-s_i^{\mathrm{tx}},y_i/f^{\mathrm{edge}}_{m,t}\},\nonumber\\
 x_i&\leftarrow x_i-r_{m,t}s_i^{\mathrm{tx}},\nonumber\\
 y_i&\leftarrow y_i-f^{\mathrm{edge}}_{m,t}s_i^{\mathrm{cp}}.
 \label{eq:two_phase_service}
\end{align}
A phase with zero residual work takes zero time. An unfinished transmission with zero link rate blocks computation, and zero CPU frequency stalls unfinished computation. The simulator uses remaining subslot capacity on subsequent tasks. Unfinished work persists across slots until completion or expiry. At each subslot, the simulator clears expired work, serves the queues, and removes unfinished tasks whose deadline equals the subslot boundary. Completion at $T_i^{\mathrm{end}}\leq T_i^{\mathrm{due}}$ contributes $D_i$ on-time bits and delay $T_i^{\mathrm{end}}-T_i^{\mathrm{arr}}$. Expired tasks contribute violated bits, including those of inactive EUs.

\subsection{Energy and Objective}
Let $\Delta\mathbf p_{u,t}$ denote the displacement used to compute propulsion energy. Rotary-wing propulsion power~\cite{zeng2019energy} for $v_{u,t}=\|\Delta\mathbf p_{u,t}\|_2/\Delta$ is
\begin{align}
 P^{\mathrm{prop}}(v)={}&P_0\left(1+\frac{3v^2}{U_{\mathrm{tip}}^2}\right)\nonumber\\
 &+P_i\left(\sqrt{\sqrt{1+\frac{v^4}{4V_0^4}}-\frac{v^2}{2V_0^2}}\right)\nonumber\\
 &+\tfrac12d_0\rho_as_{\mathrm{rot}}A_{\mathrm{disk}}v^3.
 \label{eq:propulsion_power}
\end{align}
The constants are $(P_0,P_i)=(79.86,88.63)$ W, $U_{\mathrm{tip}}=120$ m/s, $V_0=4.03$ m/s, $d_0=0.6$, $\rho_a=1.225$ kg/m$^3$, $s_{\mathrm{rot}}=0.05$, and $A_{\mathrm{disk}}=0.503$ m$^2$. Let $e_{u,t}$ be the fraction of battery capacity remaining. With radio and cubic DVFS power~\cite{wang2016mobileedge},
\begin{align}
 E_{u,t}&=\Delta\bigl[\chi_{u,t}(P^{\mathrm{prop}}(v_{u,t})+P_{\mathrm{tx}})
 +\kappa f_{u,t}^3\bigr],\nonumber\\
 E_t&=\sum_{u\in\mathcal K}E_{u,t},\nonumber\\
 e_{u,t+1}&=\max\{0,e_{u,t}-E_{u,t}/E_{\mathrm{batt}}\},
 \label{eq:battery_update}
\end{align}
Here $\kappa=10^{-28}$ W\,s$^3$, $E_{\mathrm{batt}}=548$ Wh $=1.9728$ MJ, and $\kappa F_{\max}^3=100$ W. The battery model charges every online UAV $P_{\mathrm{tx}}\Delta$ for radio use. EU uplink and local-computing energy and UAV result downlink are excluded. At $e_{u,t+1}\leq0.15$, an airborne UAV returns to the nearest of four stations and remains online in transit. It then charges offline for six slots and resumes service fully charged. Stations have unlimited concurrency, and the simulator omits collision avoidance.

For on-time bits $I_t$ and expired bits $V_t$, the reward and control objective are
\begin{align}
 r_t&=\frac{I_t-\beta V_t}{B_{\mathrm{ref}}}
 -\alpha\frac{E_t}{E_{\mathrm{hover}}K},\nonumber\\
 \max_\pi\ &\mathbb E_\pi\!\left[\sum_{t=0}^{T-1}\gamma^tr_t\right],
 \label{eq:reward_objective}
\end{align}
subject to the dynamics and bounds above. Here $\alpha=\beta=1$, $E_{\mathrm{hover}}=\Delta(P_0+P_i)$ is per-UAV hover energy per slot, and $B_{\mathrm{ref}}$ is the training 99th percentile of expected arrival bits per slot. The reward penalizes deadline violations while allowing them in feasible trajectories.

\section{RTWM World Model}
\label{sec:rtwm_world_model}

\subsection{Factorized State and Prediction}
\label{sec:world_model}
RTWM separates demand and user forecasts from action-dependent dynamics (Fig.~\ref{fig:framework}). Candidate sequences share an exogenous rollout from the same filtered state. Each candidate maintains its own UAV positions, batteries, charging modes and timers, and queue backlogs. The online posterior can depend on past actions through observed service and physical state.

\begin{figure}[t]
\centering
\includegraphics[width=1.0\columnwidth]{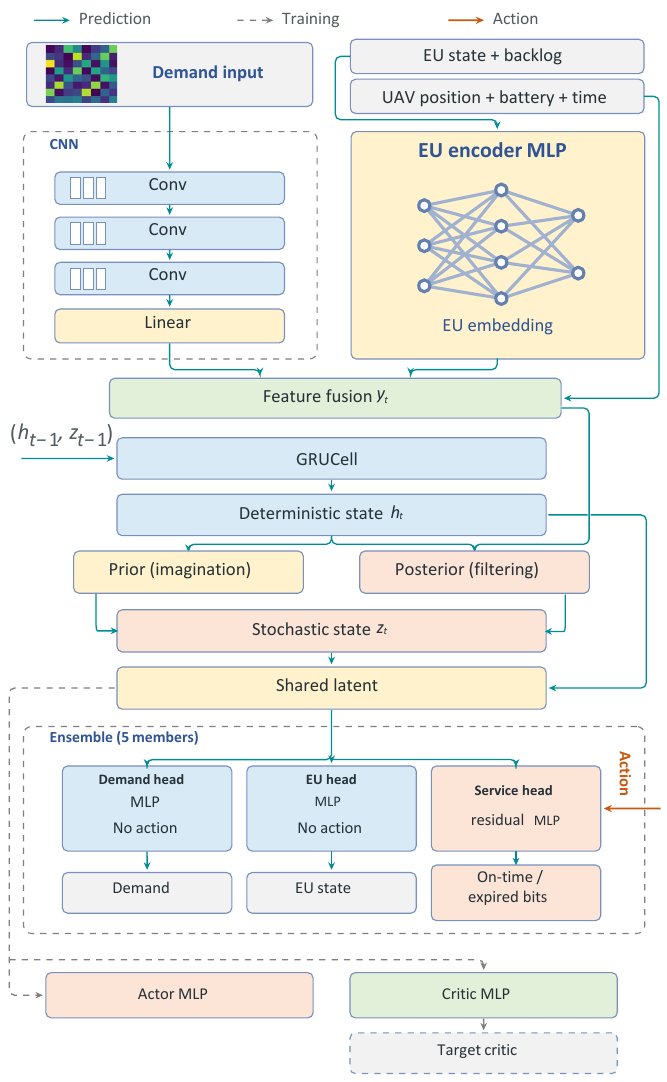}
\caption{Neural-network topology of RTWM. The recurrent latent state feeds three ensemble heads; only the service head receives the candidate action. Actor and critic share the same state input during learning.}
\label{fig:network_architecture}
\end{figure}

\subsubsection{Observation encoding and recurrent filtering}
Let $\mathbf X_t=(\mathbf x_t,\boldsymbol\zeta_t,\mathbf w_t)$ collect EU positions, activities, and demand shares. The encoder combines a demand window $\mathcal D_t$, EU fields, backlogs $\mathbf b_t$, UAV positions $\mathbf p_t$, normalized battery levels $\mathbf e_t$, and time features $\mathbf g_t$:
\begin{align}
 \mathbf y_t^{D}&=f_{\mathrm{CNN}}(\symlog(\mathcal D_t)),\nonumber\\
 \mathbf y_t^{X}&=f_{\mathrm{EU}}(\mathbf X_t,\symlog(\mathbf b_t)),\nonumber\\
 \mathbf y_t&=f_{\mathrm{enc}}(\mathbf y_t^{D},\mathbf y_t^{X},
                       \mathbf p_t,\mathbf e_t,\mathbf g_t).
 \label{eq:wm_observation_encoder}
\end{align}
Fig.~\ref{fig:network_architecture} summarizes the network widths. For nominal $K=16$, $M=32$, $G=20$, and $w=6$, the demand CNN follows $6\times20\times20\rightarrow\mathrm{Conv}_{4\times4}/2(32)\rightarrow\mathrm{Conv}_{4\times4}/2(64)\rightarrow\mathrm{Conv}_{5\times5}(128)\rightarrow\mathrm{Linear}(128)$. YJMob's $9\times9$ fields are zero-padded to $20\times20$ before this CNN. The EU encoder is an MLP with widths $5M=160\rightarrow256\rightarrow64$, and the fusion MLP has widths $128+64+3K+2=242\rightarrow256\rightarrow256$. All MLP hidden layers use ReLU and all output layers are linear. These numerical widths are nominal; the input and output dimensions below vary with $K$, $M$, and $G$.
Here $\symlog(x)=\operatorname{sgn}(x)\log(1+|x|)$ acts elementwise, and $\symexp$ is its inverse. The deterministic state $\mathbf h_t$ and stochastic state $\mathbf z_t$ satisfy
\begin{align}
 \mathbf h_t&=f_{\mathrm{GRU}}(\mathbf h_{t-1},\mathbf z_{t-1}),\nonumber\\
 p_\theta(\mathbf z_t\mid\mathbf h_t)
   &=\mathcal N(\boldsymbol\mu_\theta,
      \operatorname{diag}(\boldsymbol\sigma_\theta^2)),\nonumber\\
 q_\varphi(\mathbf z_t\mid\mathbf h_t,\mathbf y_t)
   &=\mathcal N(\boldsymbol\mu_\varphi,
      \operatorname{diag}(\boldsymbol\sigma_\varphi^2)).
 \label{eq:wm_latent_filter}
\end{align}
The recurrent filter is a GRUCell with input dimension 32 and hidden size 256. The prior MLP has widths $256\rightarrow256\rightarrow64$, and the posterior MLP has widths $512\rightarrow256\rightarrow64$; each 64-dimensional linear output parameterizes the mean and log standard deviation of a 32-dimensional diagonal Gaussian. The prior parameters depend on $\mathbf h_t$, and the posterior parameters depend on $(\mathbf h_t,\mathbf y_t)$. Training samples the posterior when filtering and the prior during imagination. Online filtering and planning use the respective means. The GRU and prior take no action input.

\subsubsection{Action-free exogenous ensembles}
Separate ensembles with $J_{\mathrm e}=5$ members decode demand and EU fields:
\begin{align}
 (\widehat D_{t+1}^{(j)},\widehat{\mathbf X}_{t+1}^{(j)})
 &=g_{\mathrm{exo}}^{(j)}(\mathbf h_t,\mathbf z_t),\nonumber\\
 \widehat{\mathbf X}_{t+1}^{(j)}
 &=(\widehat{\mathbf x}_{t+1}^{(j)},
    \widehat{\boldsymbol\zeta}_{t+1}^{(j)},
    \widehat{\mathbf w}_{t+1}^{(j)}).
 \label{eq:wm_exogenous_heads}
\end{align}
Each member receives the 288-dimensional concatenation $(\mathbf h_t,\mathbf z_t)$. Its demand head is an MLP with widths $288\rightarrow256\rightarrow G^2$ (400 nominally), and its EU head has widths $288\rightarrow256\rightarrow4M$ (128 nominally); neither head receives an action. Activity outputs are logits. Each rollout uses the ensemble means and records variance across members.

Let $\mathbf d_t=\operatorname{vec}(D_t)$ and let $\bar{\mathbf d}_t$ be the observed field at the first step and the preceding ensemble mean thereafter. Demand member $j$ predicts a residual around scaled persistence:
\begin{align}
 \mathbf l_{t+1}^{(j)}&=\rho_{\mathrm{dec}}\symlog(\bar{\mathbf d}_t)
     +\sigma_{\mathrm{dec}}\tanh g_D^{(j)}(\mathbf h_t,\mathbf z_t),\nonumber\\
 \widehat{\mathbf d}_{t+1}^{(j)}&=\pospart{\symexp(\mathbf l_{t+1}^{(j)})}.
 \label{eq:wm_residual_decode}
\end{align}
Here $\rho_{\mathrm{dec}}=0.5$, $\sigma_{\mathrm{dec}}=0.003$, and $[x]_+=\max(0,x)$ elementwise. All reported RTWM main runs explicitly enable residual decoding with these settings. Zero-initialized residual outputs start at scaled persistence. The demand loss $\mathcal L_{\mathrm{dem}}$ is ensemble mean squared error (MSE) against the residual target $\symlog(\mathbf d_{t+1})-\rho_{\mathrm{dec}}\symlog(\mathbf d_t)$. The EU loss $\mathcal L_{\mathrm{EU}}$ combines MSE for active positions and demand shares with binary cross-entropy for activities. The position loss masks inactive EUs.

\subsubsection{Action-conditioned service and queue propagation}
The physical model propagates UAV motion, service rates, battery levels, and charging state using the rules in Section~\ref{sec:system_model}. With mean task size $\bar D$ and compute density $\bar c$, it approximates serial offload service and slot arrivals by
\begin{align}
 \widetilde r_{m,t}&=\left(\frac{1}{r_{m,t}}+
                  \frac{\bar c}{f^{\mathrm{edge}}_{m,t}}\right)^{-1},\nonumber\\
 r_{\mathrm{loc}}&=\frac{f_{\mathrm{loc}}}{\bar c},\nonumber\\
 A_{m,t}&=\Delta\bar D\lambda^\star S_t w_{m,t}.
 \label{eq:wm_service_rates}
\end{align}
Here $A_{m,t}$ is in bits, and $S_t$ and $w_{m,t}$ are the rollout fields. In the nominal regime ($\xi=1$), this is the mean arrival volume from \eqref{eq:task_arrivals}. The effective offload rate is zero if either phase has zero capacity. The effective offloading ratio equals $o_{m,t}$ when a usable link exists and zero otherwise. This ratio splits both aggregate arrivals and current backlog into virtual local and offload channels. For an active EU,
\begin{align}
 W_{m,t}^{\mathrm{off}}&=\min\{b_{m,t}^{\mathrm{off}}+A_{m,t}^{\mathrm{off}},
                        \Delta\widetilde r_{m,t}\},\nonumber\\
 W_{m,t}^{\mathrm{loc}}&=\min\{b_{m,t}^{\mathrm{loc}}+A_{m,t}^{\mathrm{loc}},
                        \Delta r_{\mathrm{loc}}\},\nonumber\\
 b_{m,t+1}&=\pospart{b_{m,t}+A_{m,t}
                       -W_{m,t}^{\mathrm{off}}-W_{m,t}^{\mathrm{loc}}}.
 \label{eq:wm_queue_conservation}
\end{align}
Inactive-EU backlogs are held fixed. This fluid update omits task ordering, age, expiry, and prior queue assignment. Resource shares follow demand weights, so capacity reserved for an EU can remain unused.

A five-member residual ensemble receives the latent state, candidate action, and symlog-transformed service, backlog, and arrival summaries. Each member is an MLP with widths $288+(4K+M)+3M=480\rightarrow256\rightarrow2$ in the nominal configuration. After symexp and nonnegative clipping, its two outputs directly represent the slot-level totals $\widehat I_t^{(j)}$ and $\widehat V_t^{(j)}$, rather than separate predictions for each EU. The service loss $\mathcal L_{\mathrm{srv}}$ compares the symlog of each ensemble-mean total with its observed counterpart $I_t$ or $V_t$ by MSE. These residuals do not update the fluid backlog, and their predicted outcomes need not satisfy queue conservation or service-capacity bounds.

\subsubsection{Multi-step training}
The world-model loss is
\begin{align}
 \mathcal L_{\mathrm{WM}}={}&\mathcal L_{\mathrm{dem}}
 +\mathcal L_{\mathrm{EU}}+0.5\mathcal L_{\mathrm{exo}}^{\mathrm{multi}}
 \nonumber\\
 &+\mathcal L_{\mathrm{srv}}+\mathcal L_q+\mathcal L_{\mathrm{KL}},
 \label{eq:wm_total_loss}
\end{align}
where $\mathcal L_{\mathrm{exo}}^{\mathrm{multi}}$ recursively predicts demand and EU fields through the prior over $H_x=4$ steps. The balanced free-bits KL term uses a threshold of 1 nat, a balance weight of 0.8, and complementary stop-gradients on the prior and posterior. The queue contribution is
\begin{equation}
 \mathcal L_q=\mathcal L_q^{(1)}
              +\frac{0.5}{H_q}\sum_{k=1}^{H_q}\mathcal L_q^{(k)},
 \qquad H_q=4,
 \label{eq:queue_loss}
\end{equation}
where $\mathcal L_q^{(k)}$ is symlog MSE against the replay backlog at horizon $k$. All steps use recorded actions and UAV positions, battery levels, modes, and timers. Backlog rolls forward recursively, and exogenous fields are decoded recursively after the first step. The one-step term has zero model gradient because it uses only recorded inputs and fixed rules. Terms at $k\geq2$ train the model through its exogenous predictions.

\subsection{Imagination-Based Learning and Decision-Time Planning}
\label{sec:planning}

The actor and critic both consume $(\mathbf h_t,\mathbf z_t,\mathbf e_t,\symlog(\mathbf b_t))$, of dimension $256+32+K+M=336$ nominally. The actor MLP has widths $336\rightarrow256\rightarrow256\rightarrow192$, and the critic MLP has widths $336\rightarrow256\rightarrow256\rightarrow1$. The actor's linear output contains the Gaussian mean and log standard deviation for each of the $4K+M=96$ action coordinates, followed by the bounded sampling below.
Training starts imagination from replay posterior samples and observed physical states. The actor samples bounded actions as
\begin{align}
 (\boldsymbol\mu_{\vartheta,t},\boldsymbol\sigma_{\vartheta,t})
   &=\pi_\vartheta(\mathbf h_t,\mathbf z_t,\mathbf e_t,
                   \symlog(\mathbf b_t)),\nonumber\\
 \mathbf a_t&=\tanh(\boldsymbol\mu_{\vartheta,t}
            +\boldsymbol\sigma_{\vartheta,t}\odot\boldsymbol\epsilon_t),
 \nonumber\\
 \boldsymbol\epsilon_t&\sim\mathcal N(\mathbf 0,\mathbf I).
 \label{eq:actor_sample}
\end{align}
Training and planning both use $H=12$ slots. Within a rollout, index $k=0,\ldots,H-1$ denotes the offset from its initial state. The service model evaluates each action and advances backlog, motion, batteries, and charging. The ensemble-mean outcomes and predicted fleet energy give $\widetilde r_k$ via \eqref{eq:reward_objective}. The exogenous decoder and latent prior then advance. Actor rollouts sample the prior, and CEM uses its mean.

To quantify ensemble disagreement, let $I_k^{(j)}=\widehat I_k^{(j)}$ and $V_k^{(j)}=\widehat V_k^{(j)}$ be the residual members' slot-level on-time and expired-bit totals. Then
\begin{align}
 u_k^{\mathrm{res}}&=\Var_j(I_k^{(j)}/B_{\mathrm{ref}})
                     +\Var_j(V_k^{(j)}/B_{\mathrm{ref}}),\nonumber\\
 u_k^{\mathrm{exo}}&=\operatorname{mean}_g\Var_j(\widehat d_{k+1,g}^{(j)})
  +\operatorname{mean}_q\Var_j(\widehat X_{k+1,q}^{\mathrm{raw},(j)}).
 \label{eq:disagreement}
\end{align}
Demand disagreement uses the nonnegative decoded field. EU disagreement averages over raw positions, activity logits, and demand shares; $q$ indexes their concatenated coordinates. The exogenous terms retain their respective scales. Training uses $u_k^{\mathrm{AC}}=u_k^{\mathrm{res}}+u_k^{\mathrm{exo}}$ and includes an exogenous decode after the last reward. In CEM, $u_k^{\mathrm{CEM}}=u_k^{\mathrm{res}}+\mathbb I\{k+1<H\}u_k^{\mathrm{exo}}$. The exogenous penalty is constant across candidates from the same initial state and cancels in their ranking.

With $\gamma=0.997$ and $\lambda_{\mathrm{ret}}=0.95$, the critic trains on imagined returns
\begin{align}
 G_H&=\bar v_H,\nonumber\\
 G_k&=\widetilde r_k-u_k^{\mathrm{AC}}
 +\gamma[(1-\lambda_{\mathrm{ret}})\bar v_{k+1}
                   +\lambda_{\mathrm{ret}} G_{k+1}],
 \label{eq:lambda_return}
\end{align}
where $\bar v_k$ is the target-critic output after symexp and clipping to $[-Hr_{\lim},Hr_{\lim}]$. Imagined rewards are clipped to $[-r_{\lim},r_{\lim}]$, with $r_{\lim}=1.1[(1+\beta)+1.7\alpha]$. The critic predicts $\symlog(G_k)$ from detached imagined states. The target uses a Polyak update rate of 0.005. The actor objective uses the immediate imagined reward minus disagreement and adds pre-tanh Gaussian entropy with weight $10^{-3}$. Gradients through the target values and recursive return branch are stopped. The actor is used for training and data collection.

\begin{figure}[!htb]
\centering
\includegraphics[width=1.0\columnwidth]{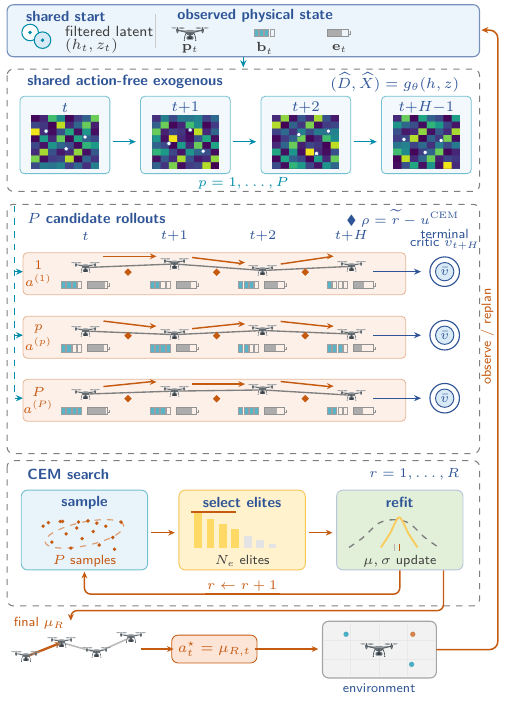}
\caption{Candidate scoring and receding-horizon CEM. The controller refits elites across $R$ rounds, executes the first action of the final mean, and replans from the next observation.}
\label{fig:rtwm_rollout}
\end{figure}

At deployment, all candidates start from the current posterior mean and observed physical state (Fig.~\ref{fig:rtwm_rollout}). CEM scores each bounded sequence $\mathbf a_{0:H-1}$ by
\begin{equation}
 \mathcal J(\mathbf a_{0:H-1})
 =\sum_{k=0}^{H-1}\gamma^k(\widetilde r_k-u_k^{\mathrm{CEM}})
   +\gamma^H\bar v_H.
 \label{eq:cem_score}
\end{equation}
The first service step uses observed demand and EU fields. Later steps use decoded spatial and EU fields while holding aggregate arrival intensity at its initial value ($\rho_{\mathrm{plan}}=1$). This planning rule is separate from the decoder's persistence coefficient $\rho_{\mathrm{dec}}$.

Each decision resets a diagonal Gaussian to zero mean and standard deviation 0.6. Each of $R=5$ rounds draws $P=512$ action sequences and clips them to $[-1,1]$. A fixed set of relocation sequences replaces the first $N_{\mathrm{ray}}$ samples:
\begin{align}
 N_{\mathrm{ray}}&=\begin{cases}
 1+24K,&1+24K\leq P,\\
 1+12K,&\text{balanced overflow},
 \end{cases}\nonumber\\
 N_{\mathrm{Gaussian}}&=P-N_{\mathrm{ray}}.
 \label{eq:ray_counts}
\end{align}
Each relocation sequence holds a direction and speed over the horizon and uses equal bandwidth logits, full CPU frequency, and full offloading. At $K=16$, 385 sequences are fixed and 127 are Gaussian draws. At $K=32$, the 769-ray set overflows the 512-candidate pool; balanced truncation keeps hover plus 384 rays spread over all three speed tiers and every UAV direction, leaving 127 Gaussian draws. Order-retaining truncation instead keeps the first 512 rays and no Gaussian draw, and is not used for the reported $K=32$ arm. The planner refits the coordinate-wise mean and standard deviation to the 32 elites, with a standard deviation floor of $10^{-3}$. It executes the first action of the final mean and replans after filtering the next observation. Algorithm~\ref{alg:rtwm} summarizes the procedure.

\begin{figure}[!htb]
\centering
\begin{minipage}{\columnwidth}
\hrule\smallskip
\refstepcounter{rtwmalg}\label{alg:rtwm}
\textbf{Algorithm \thertwmalg.} RTWM training and control
\smallskip\hrule\smallskip
\begin{algorithmic}[1]
\STATE \textbf{Training, each update:}
\STATE Filter replay observations and minimize \eqref{eq:wm_total_loss}.
\STATE Roll out $H$ steps with actor actions and the action-free prior.
\STATE Fit the critic to $\symlog(G_k)$ and update its target.
\STATE Update the actor from immediate adjusted rewards and entropy.
\STATE \textbf{Deployment, each slot:}
\STATE Filter the observation and synchronize physical states.
\STATE Reset the Gaussian and construct the relocation sequences.
\FOR{$r=1$ to $R$}
 \STATE Draw and clip $P$ sequences. Replace the first $N_{\mathrm{ray}}$.
 \STATE Score by \eqref{eq:cem_score} and refit to the top 32.
\ENDFOR
\STATE Execute the first action of the final mean.
\end{algorithmic}
\smallskip\hrule
\end{minipage}
\end{figure}

\section{Experimental Results}
\label{sec:experimental_results}

\subsection{Experimental Setup}
\label{sec:exp_datasets}
Table~\ref{tab:exp_datasets} summarizes the traces and chronological splits. Milan uses Internet-CDR counts~\cite{barlacchi2015multisource}, Shanghai uses session initiations~\cite{mexwell2023telecom,li2022profit,guo2020user,wang2021delay}, and YJMob uses counts of distinct users present~\cite{yabe2024yjmob100k}, all replayed at 10-min resolution. Shanghai applies a three-slot trailing average, and YJMob divides each 30-min count uniformly across three slots. Zero demand produces no arrivals. Training data determine the 99th-percentile scale, cross-dataset intensity alignment, EU count calibration, $\lambda^\star$, and $B_{\mathrm{ref}}$.

Each seed uses three regular and four secondary evaluation days. Milan's regular group includes calendar weekend days. Captions use \emph{regular} for plots labeled \emph{Weekday} and \emph{secondary} for plots labeled \emph{Holiday} or \emph{Holiday/weekend}, denoting Milan holidays, Shanghai weekends, or YJMob inferred pseudo-weekends. Shanghai's crop window, grid alignment, and three-slot average are fitted on training days only, so smoothing never crosses the train/test boundary. YJMob excludes day 27 as the source recommends, and its crop window and pseudo-weekend labels are inferred from training days only. All Shanghai and YJMob main-table results use this train-only preprocessing and calibration protocol. The top 1\% of cells account for 4.4\%, 8.0\%, and 9.3\% of demand at the peak training hour on Milan, Shanghai, and YJMob, respectively.
\begin{table*}[t]
\centering
\caption{Chronological splits and fixed evaluation days. The regular partition is primary.}
\label{tab:exp_datasets}
\footnotesize
\setlength{\tabcolsep}{2.5pt}
\begin{tabular}{@{}lp{2.8cm}p{2.3cm}p{2.3cm}p{2.7cm}p{3.4cm}@{}}
\toprule
Dataset & Demand proxy and grid & Training days & Test days & Regular evaluation & Secondary evaluation \\
\midrule
Milan & Internet-CDR count; $20\times20$, 235-m cells & 1--10 Nov. 2013 (10 d) & 21 Dec. 2013--1 Jan. 2014 (12 d) & 21--23 Dec. 2013 & 24--25 Dec. 2013; 31 Dec. 2013; 1 Jan. 2014 \\
Shanghai & Session initiations; $20\times20$, 470-m cells & 31 May--31 Oct. 2014 (154 d) & 1--30 Nov. 2014 (30 d) & 3--5 Nov. 2014 & 1--2 and 8--9 Nov. 2014 \\
YJMob & Distinct-user presence; $9\times9$, 500-m cells & Days 0--55, excl. 27 (55 d) & Days 56--74 (19 d) & Days 57--59 & Days 62, 63, 69, and 70 \\
\bottomrule
\end{tabular}
\end{table*}

Eight offline EU realizations per dataset are sampled at episode reset independently of UAV actions. For aggregate demand $S_t^{\mathrm{EU}}$, the active count is
\begin{equation}
 N_t=\min\{M,\operatorname{round}(\kappa_{\mathrm{EU}}S_t^{\mathrm{EU}})\},
\end{equation}
where $\kappa_{\mathrm{EU}}$ is fitted on training data. The materializer allocates users by cell demand with fixed within-cell offsets, retains each user's cell when possible, and otherwise assigns the nearest cell with an unmet count. Active-EU shares recover aggregate arrival intensity when $N_t>0$. The nominal setting uses $K=16$ UAVs, $M=32$ EUs, a six-slot demand window, and one-day episodes. Evaluation replays test-day exogenous fields without gradient updates.

RTWM uses a 128-dimensional demand embedding, a 256-dimensional recurrent hidden state, a 32-dimensional stochastic state, and five-member ensembles (Fig.~\ref{fig:network_architecture}). The world-model and actor--critic parameters use Adam with learning rate $3\times10^{-4}$. Each update samples 32 replay sequences of length 32 from a 60{,}000-transition replay buffer and clips gradient norms at 100. We set $H_x=H_q=4$ for the multi-step losses, $H=12$ for imagination and planning, and $\tau=0.005$ for the target critic. The production operating point collects 40{,}000 interactions and performs 8{,}000 gradient updates (Table~\ref{tab:training_planning_budgets}).

All methods share the simulator, observations, and feasible actions. The baselines are Static, Greedy, INS-WOA, PPO, SAC, MADDPG, GA-MATR, and a compact TD-MPC-style latent-CEM comparator with a deterministic latent model, terminal target value, shifted mean, and no actor or policy prior. Greedy and INS-WOA optimize a coverage surrogate; RTWM and the learning methods use the deadline-and-energy reward.
\begin{table}[!htbp]
\centering
\caption{Method-specific budgets. Imagined transitions count candidates across all horizons and CEM rounds before the ensemble cost.}
\label{tab:training_planning_budgets}
\footnotesize
\setlength{\tabcolsep}{3pt}
\begin{tabular}{@{}lcc@{}}
\toprule
Budget & RTWM & TD-MPC-style \\
\midrule
Collection steps & 40{,}000 & 5{,}000 \\
Gradient/model updates & 8{,}000 & 50{,}000 \\
Horizon $H$ & 12 & 12 \\
Candidates $\times$ CEM rounds & $512\times5$ & $256\times3$ \\
Transitions per decision & 30{,}720 & 9{,}216 \\
Rays / CEM initialization & 385 / reset mean & 0 / mean shift \\
Discount $\gamma$ & 0.997 & 0.99 \\
\bottomrule
\end{tabular}
\end{table}

Each RTWM decision scores $RPH=30{,}720$ transitions before batching, with $O(RPHJ_{\mathrm e})$ ensemble evaluations plus association and resource costs. PPO and GA-MATR use 300{,}000 collection steps, MADDPG uses 100{,}000, and SAC uses 200{,}000; Static, Greedy, and INS-WOA require no gradient training. The production experiments therefore compare complete implementations under unequal budgets (Table~\ref{tab:training_planning_budgets}).

\subsection{Metrics and Statistical Analysis}
\label{sec:exp_metrics}
The primary metric is $\mathrm{DSR}=N_{\mathrm{in}}/N_{\mathrm{arr}}$, with $N_{\mathrm{in}}$ on-time tasks and $N_{\mathrm{arr}}$ arrivals. Secondary metrics include bit-weighted DSR, completion and deadline violation ratios, mean and 90th-percentile delay, fleet energy (kJ/slot), and aerial-fleet energy efficiency (on-time Mbit/J). Delay statistics use only on-time completions. We compute ratios per day and average daily metrics equally within each calendar group. Unfinished tasks at episode end count only in the arrival denominator, so DSR and violation need not sum to one. The maximum unresolved fraction in the archived main runs is approximately 0.033\%.

The analysis includes archived completed finite runs: ten per dataset for RTWM and the TD-MPC-style comparator, and ten Milan runs plus ten Shanghai and YJMob runs for each other baseline. For each shared seed $s$, we compute $\Delta_s=x_s^{\mathrm{RTWM}}-x_s^{\mathrm{base}}$ from the day-group means. The nominal two-sided 95\% seed-based paired interval is $\overline\Delta\pm t_{0.975,n-1}s_\Delta/\sqrt n$, where $s_\Delta$ is the sample SD across $n$ paired differences. 

\begin{figure*}[t]
\centering
\includegraphics[width=0.8\textwidth]{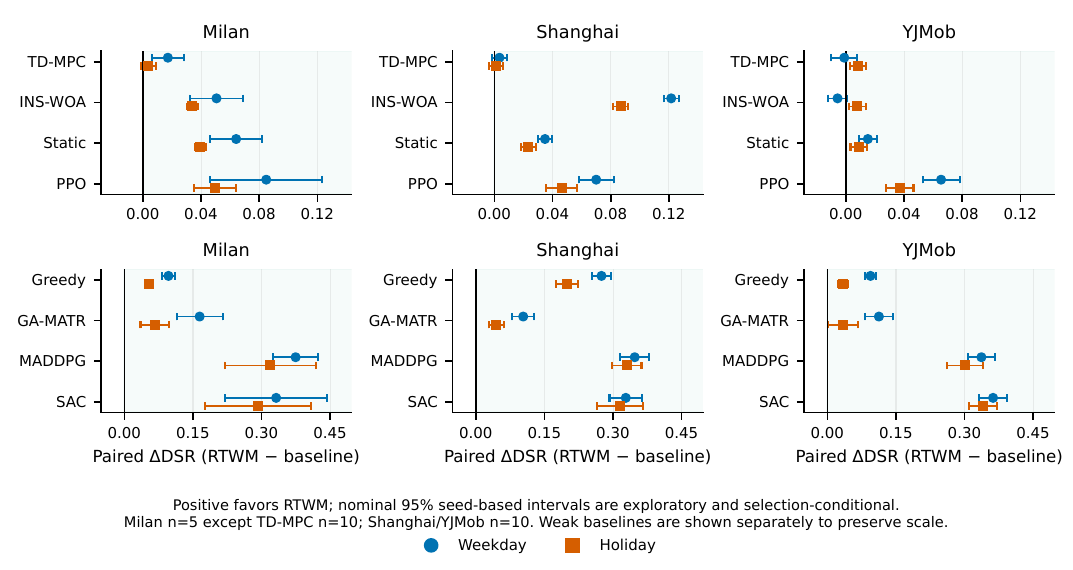}
\caption{Paired DSR differences (RTWM minus baseline). Table~\ref{tab:main_dsr} reports absolute DSR. Milan comparisons against baselines other than TD-MPC-style use $n=5$; all others use $n=10$. Paired mean differences cannot be directly recomputed from the ten-seed RTWM marginal means in Table~\ref{tab:main_dsr}. Bars show nominal two-sided 95\% seed-based intervals for training-seed uncertainty on the fixed reused test days, not significance tests. The upper and lower panels use different horizontal scales.}
\label{fig:paired_dsr}
\end{figure*}

\subsection{Deadline Satisfaction}
Table~\ref{tab:main_dsr} reports the primary DSR comparison under method-specific, unequal budgets. On regular days, RTWM's paired mean DSR differences from the TD-MPC-style comparator are $+0.0172$ on Milan, $+0.0035$ on Shanghai, and $-0.0011$ on YJMob (Fig.~\ref{fig:paired_dsr}), the last a descriptive 0.1-percentage-point deficit. RTWM attains the highest observed regular-day mean DSR on Milan and Shanghai. On YJMob, it trails only the INS-WOA heuristic among the non-learned references and leads Static by 1.5 percentage points.
\begin{table*}[t]
\centering
\caption{DSR under method-specific budgets (mean $\pm$ sample SD; $n=10$ except Milan $\dagger$, $n=5$). Regular/Secondary follow Table~\ref{tab:exp_datasets}. Greedy and INS-WOA use a coverage surrogate. All comparisons are exploratory and selection-conditional.}
\label{tab:main_dsr}
\footnotesize
\setlength{\tabcolsep}{3pt}
\begin{tabular}{lcccccc}
\toprule
Method & \multicolumn{2}{c}{Milan} & \multicolumn{2}{c}{Shanghai} & \multicolumn{2}{c}{YJMob} \\
\cmidrule(lr){2-3}\cmidrule(lr){4-5}\cmidrule(lr){6-7}
 & Regular & Secondary & Regular & Secondary & Regular & Secondary \\
\midrule
\textbf{RTWM} & $0.9522\,\pm\,0.0138$ & $0.9665\,\pm\,0.0075$ & $0.9287\,\pm\,0.0071$ & $0.9584\,\pm\,0.0073$ & $0.8652\,\pm\,0.0086$ & $0.9025\,\pm\,0.0079$ \\
TD-MPC-style & $0.9350\,\pm\,0.0082$ & $0.9630\,\pm\,0.0032$ & $0.9252\,\pm\,0.0037$ & $0.9572\,\pm\,0.0028$ & $0.8663\,\pm\,0.0100$ & $0.8939\,\pm\,0.0029$ \\
INS-WOA & $0.9100\,\pm\,0.0014$$^{\dagger}$ & $0.9385\,\pm\,0.0007$$^{\dagger}$ & $0.8071\,\pm\,0.0009$ & $0.8714\,\pm\,0.0011$ & $0.8709\,\pm\,0.0024$ & $0.8945\,\pm\,0.0008$ \\
Static & $0.8965\,\pm\,0.0009$$^{\dagger}$ & $0.9331\,\pm\,0.0009$$^{\dagger}$ & $0.8939\,\pm\,0.0008$ & $0.9353\,\pm\,0.0006$ & $0.8501\,\pm\,0.0053$ & $0.8936\,\pm\,0.0018$ \\
PPO & $0.8758\,\pm\,0.0214$$^{\dagger}$ & $0.9232\,\pm\,0.0139$$^{\dagger}$ & $0.8587\,\pm\,0.0132$ & $0.9122\,\pm\,0.0114$ & $0.7997\,\pm\,0.0202$ & $0.8655\,\pm\,0.0092$ \\
Greedy & $0.8642\,\pm\,0.0062$$^{\dagger}$ & $0.9193\,\pm\,0.0054$$^{\dagger}$ & $0.6540\,\pm\,0.0292$ & $0.7587\,\pm\,0.0319$ & $0.7709\,\pm\,0.0188$ & $0.8679\,\pm\,0.0113$ \\
GA-MATR & $0.7961\,\pm\,0.0384$$^{\dagger}$ & $0.9059\,\pm\,0.0239$$^{\dagger}$ & $0.8252\,\pm\,0.0343$ & $0.9140\,\pm\,0.0232$ & $0.7526\,\pm\,0.0431$ & $0.8676\,\pm\,0.0412$ \\
MADDPG & $0.5861\,\pm\,0.0378$$^{\dagger}$ & $0.6534\,\pm\,0.0816$$^{\dagger}$ & $0.5813\,\pm\,0.0477$ & $0.6280\,\pm\,0.0420$ & $0.5283\,\pm\,0.0384$ & $0.6016\,\pm\,0.0586$ \\
SAC & $0.6287\,\pm\,0.0849$$^{\dagger}$ & $0.6810\,\pm\,0.0931$$^{\dagger}$ & $0.6008\,\pm\,0.0493$ & $0.6430\,\pm\,0.0672$ & $0.5028\,\pm\,0.0419$ & $0.5620\,\pm\,0.0456$ \\
\bottomrule
\end{tabular}
\end{table*}

For the secondary partition, RTWM's mean DSR is higher than the comparator's on all three datasets. The corresponding differences between the Table~\ref{tab:main_dsr} means are 0.35, 0.12, and 0.86 percentage points on Milan, Shanghai, and YJMob, respectively. Together with the regular-day results, these comparisons locate RTWM's largest DSR margin over the comparator on Milan and show consistent secondary-partition gains across all three traces. Fig.~\ref{fig:paired_dsr} reports all 48 paired intervals.

\begin{figure}[!htb]
\centering
\includegraphics[width=0.90\columnwidth]{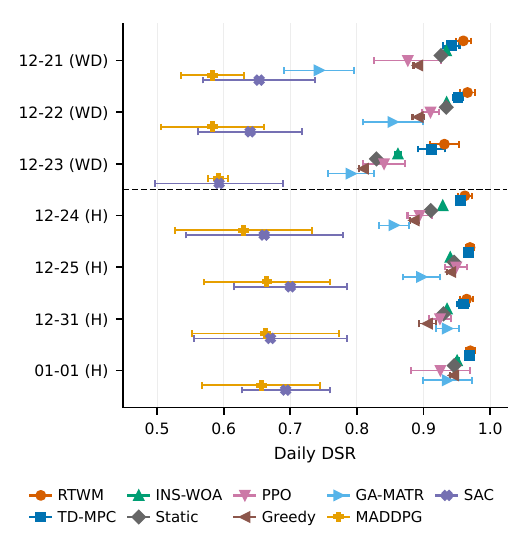}

\caption{Daily Milan DSR for the nine methods. Markers and bars show means and sample SDs. WD and H denote regular and holiday evaluation days.}
\label{fig:milan_daily}
\end{figure}

Fig.~\ref{fig:milan_daily} reports DSR for the nine methods on each Milan evaluation day. On 23 December, RTWM, the TD-MPC-style comparator, INS-WOA, and Static all have lower DSR than on 22 December. RTWM and the comparator remain close throughout the holiday partition.

\subsection{Budget-Matched Controls and Runtime}
A five-seed Milan control matches collection at 40{,}000 interactions: RTWM's regular-day DSR of 0.9382 is below TD-MPC-style's 0.9420 and PPO's 0.9394, while the secondary-day means are 0.9672, 0.9661, and 0.9766, respectively. This control therefore supports neither a sample-efficiency advantage nor a general algorithmic advantage for RTWM. The PPO main experiment at 300{,}000 interactions and the 40{,}000-interaction control are separate runs, not a controlled learning curve. A separate five-seed control gives RTWM and TD-MPC-style regular/secondary means of 0.9493/0.9677 and 0.9340/0.9624 at $256\times3$ candidates and rounds. In this planning-matched control, RTWM's mean is 1.53 percentage points higher on regular days and 0.53 points higher on secondary days. The control matches only candidate and round counts: the ray pool, CEM initialization, discount, and training choices remain unmatched, so it cannot isolate an algorithmic contribution.

Training wall-clock and decision latency were measured on the shared dual-GPU workstation under concurrent multi-job load. Training times are available for the instrumented RTWM runs across datasets and the instrumented Milan SAC run: 5.1k--17.2k~s and 10.0k~s, respectively. The per-decision p95 is 0.31--1.49~s for RTWM, 0.03--0.12~s for TD-MPC-style, and about 1~ms for Milan SAC. In the planning-matched Milan control, the regular-day p95 is 0.665~s for RTWM and 0.111~s for TD-MPC-style. These are load-conditional operating points, not idle-device or guaranteed real-time bounds. Timing measurements come from the instrumented runs identified here; the other Milan main-table runs have no timing measurements.

\subsection{Delay, Deadline Violation, and Fleet Energy}
\begin{figure}[!htb]
\centering
\includegraphics[width=.4\columnwidth,trim=75.5bp 135bp 302bp 8bp,clip]{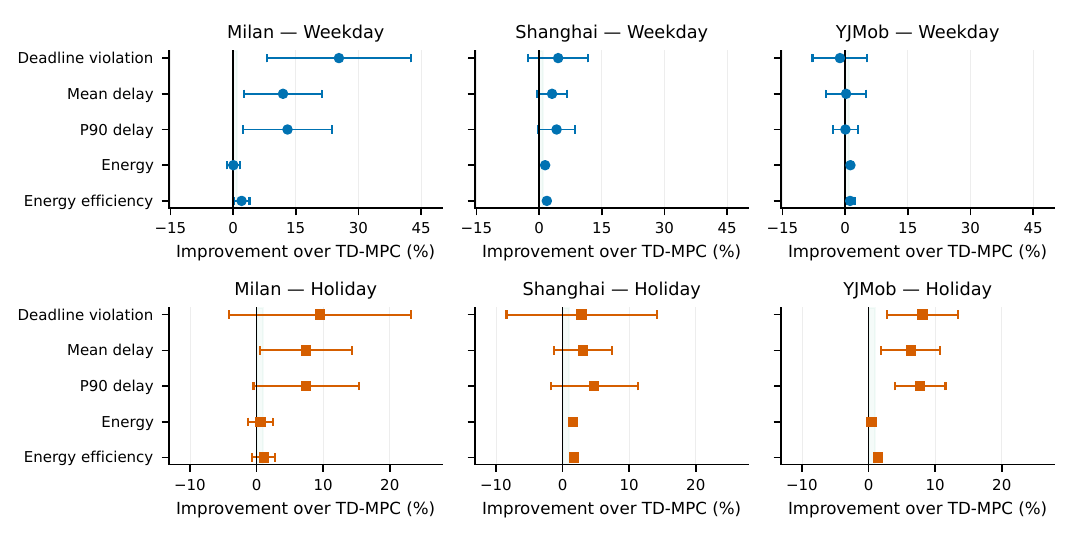}\hfill%
\includegraphics[width=.4\columnwidth,trim=75.5bp 12.5bp 302bp 131bp,clip]{results/fig13_it1_secondary.pdf}\par\smallskip
\includegraphics[width=.4\columnwidth,trim=221bp 135bp 157bp 8bp,clip]{results/fig13_it1_secondary.pdf}\hfill%
\includegraphics[width=.4\columnwidth,trim=221bp 12.5bp 157bp 131bp,clip]{results/fig13_it1_secondary.pdf}\par\smallskip
\includegraphics[width=.4\columnwidth,trim=367.5bp 135bp 10.5bp 8bp,clip]{results/fig13_it1_secondary.pdf}\hfill%
\includegraphics[width=.4\columnwidth,trim=367.5bp 12.5bp 10.5bp 131bp,clip]{results/fig13_it1_secondary.pdf}
\caption{Paired relative changes versus the TD-MPC-style comparator (ten seeds). For each seed, with $a$ denoting RTWM and $b$ the comparator, violation, delay, and energy use $100(b-a)/|b|$, while energy efficiency (EE) uses $100(a-b)/|b|$. Within each panel, rows from top to bottom show deadline violation, mean delay, P90 delay, energy, and energy efficiency. Positive changes favor RTWM. Bars show nominal two-sided 95\% seed-based intervals for training-seed uncertainty on the fixed reused test days, not significance tests.}
\label{fig:paired_secondary}
\end{figure}

Fig.~\ref{fig:paired_secondary} compares the secondary metrics with the TD-MPC-style comparator. On regular Milan days, RTWM has lower deadline violation, delay, and energy. On Shanghai, delay and energy are lower in both partitions. On regular YJMob days, RTWM has lower fleet energy and a higher violation ratio. On YJMob pseudo-weekends, RTWM has higher DSR and lower delay and energy. Thus, the DSR gains on Milan and Shanghai coincide with improvements in delay or energy, while the YJMob pseudo-weekend results combine higher DSR with lower delay and energy. The delay comparison is conditional on on-time completion. \ifextendedfigures The extended results plot absolute metrics and empirical energy--DSR trade-offs.\fi

\subsection{Demand, EU Count, and Fleet Size}
\begin{figure}[!htb]
\centering
\includegraphics[width=0.75\columnwidth]{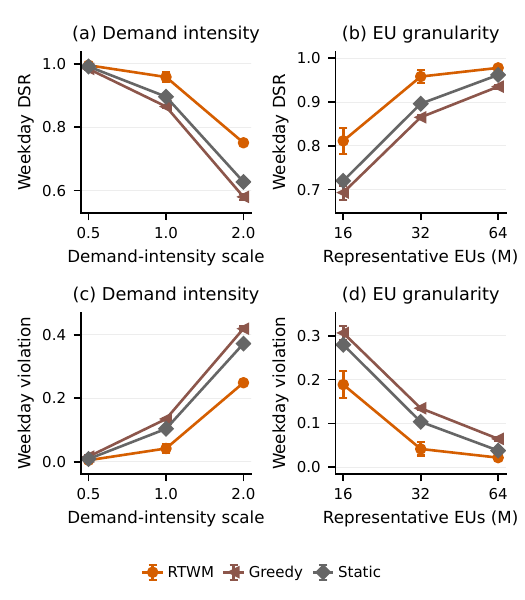}
\caption{Regular-day DSR and violation on Milan across demand multipliers $\xi$ and representative-EU counts $M$ . Table~\ref{tab:sensitivity} gives exact DSR means and sample SDs for both calendar partitions.}
\label{fig:demand_eu_sensitivity}
\end{figure}

\begin{figure}[!htb]
\centering
\includegraphics[width=0.75\columnwidth]{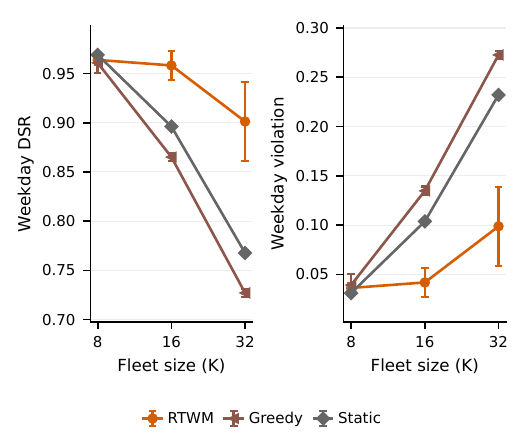}
\caption{Regular-day DSR and violation for Milan fleets of $K=8,16,32$ UAVs. Table~\ref{tab:sensitivity} gives exact DSR means and sample SDs for both calendar partitions.}
\label{fig:k_sensitivity}
\end{figure}

Figs.~\ref{fig:demand_eu_sensitivity} and~\ref{fig:k_sensitivity} show regular-day DSR and deadline violations in Milan sensitivity runs. Table~\ref{tab:sensitivity} reports DSR means and sample SDs for both calendar partitions.

At half demand, DSR approaches one for all three methods. Doubling demand lowers DSR across methods and widens RTWM's lead over Static to 12.34 percentage points. RTWM's regular-day mean changes from 0.9582 in the nominal setting to 0.7509 at doubled demand, while its margin over Static increases from 6.20 to 12.34 percentage points. RTWM and Static both improve as $M$ increases, which also changes the EU realization, calibration, aggregate local compute, and FIFO parallelism. Across $M=16,32,64$, RTWM's regular-day means are 0.8112, 0.9582, and 0.9780, respectively. At $K=32$, RTWM reaches 0.9012, exceeding Static and Greedy by 13.35 and 17.41 percentage points, respectively.

\begin{table*}[t]
\centering
\caption{Exploratory Milan DSR across joint operating regimes (mean $\pm$ sample SD, seeds 0--2, $n=3$). $M$ also changes EU realization, calibration, queues, and local compute; $K$ additionally changes arrival/reward calibration and CEM coverage.}
\label{tab:sensitivity}
\scriptsize
\setlength{\tabcolsep}{3pt}
\begin{tabular}{llcccccc}
\toprule
Factor & Setting & \multicolumn{2}{c}{RTWM} & \multicolumn{2}{c}{Static} & \multicolumn{2}{c}{Greedy} \\
\cmidrule(lr){3-4}\cmidrule(lr){5-6}\cmidrule(lr){7-8}
 & & Regular & Secondary & Regular & Secondary & Regular & Secondary \\
\midrule
Nominal & $M{=}32,K{=}16$ & $0.9582\,\pm\,0.0150$ & $0.9743\,\pm\,0.0022$ & $0.8962\,\pm\,0.0011$ & $0.9332\,\pm\,0.0010$ & $0.8651\,\pm\,0.0044$ & $0.9195\,\pm\,0.0071$ \\
\midrule
Demand & $0.5\times$ & $0.9952\,\pm\,0.0010$ & $0.9992\,\pm\,0.0002$ & $0.9913\,\pm\,0.0007$ & $0.9975\,\pm\,0.0001$ & $0.9838\,\pm\,0.0046$ & $0.9966\,\pm\,0.0008$ \\
 & $2.0\times$ & $0.7509\,\pm\,0.0085$ & $0.7987\,\pm\,0.0103$ & $0.6275\,\pm\,0.0036$ & $0.7294\,\pm\,0.0023$ & $0.5803\,\pm\,0.0085$ & $0.6693\,\pm\,0.0100$ \\
\midrule
Representative EUs & $M=16$ & $0.8112\,\pm\,0.0300$ & $0.8406\,\pm\,0.0251$ & $0.7200\,\pm\,0.0065$ & $0.7876\,\pm\,0.0019$ & $0.6931\,\pm\,0.0159$ & $0.7728\,\pm\,0.0266$ \\
 & $M=64$ & $0.9780\,\pm\,0.0066$ & $0.9910\,\pm\,0.0009$ & $0.9619\,\pm\,0.0001$ & $0.9892\,\pm\,0.0003$ & $0.9351\,\pm\,0.0043$ & $0.9766\,\pm\,0.0034$ \\
\midrule
Fleet & $K=8$ & $0.9638\,\pm\,0.0053$ & $0.9861\,\pm\,0.0042$ & $0.9689\,\pm\,0.0003$ & $0.9882\,\pm\,0.0005$ & $0.9608\,\pm\,0.0107$ & $0.9848\,\pm\,0.0019$ \\
 & $K=32$ & $0.9012\,\pm\,0.0399$ & $0.9298\,\pm\,0.0260$ & $0.7677\,\pm\,0.0025$ & $0.8288\,\pm\,0.0008$ & $0.7271\,\pm\,0.0043$ & $0.7925\,\pm\,0.0136$ \\
\bottomrule
\end{tabular}
\end{table*}

\subsection{Training Progress}
\begin{figure}[!htb]
\centering
\includegraphics[width=0.48\columnwidth,trim=0bp 8bp 337bp 7bp,clip]{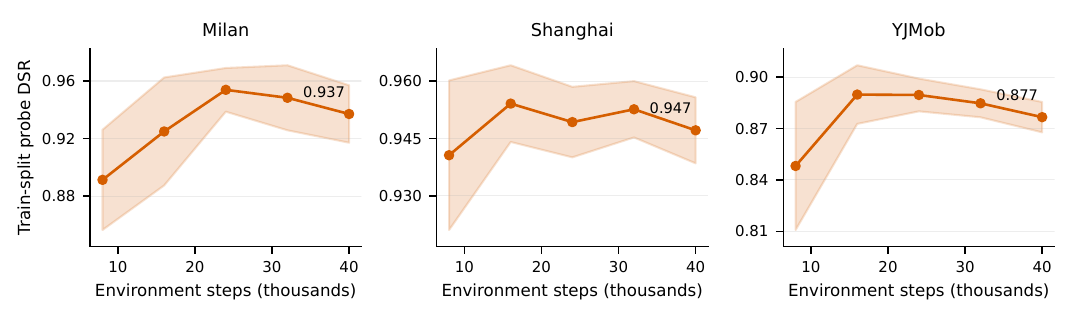}\hfill%
\includegraphics[width=0.48\columnwidth,trim=177bp 8bp 170bp 7bp,clip]{results/fig04_it1_curve.pdf}\par\smallskip
\includegraphics[width=0.48\columnwidth,trim=345bp 8bp 0bp 7bp,clip]{results/fig04_it1_curve.pdf}\par
\caption{Train-split probe DSR over ten seeds. The panels show Milan, Shanghai, and YJMob in reading order. Shading denotes sample SD.}
\label{fig:train_probe}
\end{figure}

Fig.~\ref{fig:train_probe} gives train-split DSR at five prespecified collection checkpoints. The Milan probe uses the final training day. All three datasets show substantial gains before 20{,}000 environment steps, followed by smaller changes through 40{,}000 steps. The final annotated probe means are 0.937 on Milan, 0.947 on Shanghai, and 0.877 on YJMob.

\subsection{Open-Loop Prediction on Milan}
\begin{figure}[!htb]
\centering
\includegraphics[width=0.48\columnwidth,trim=0bp 21bp 343.2bp 3bp,clip]{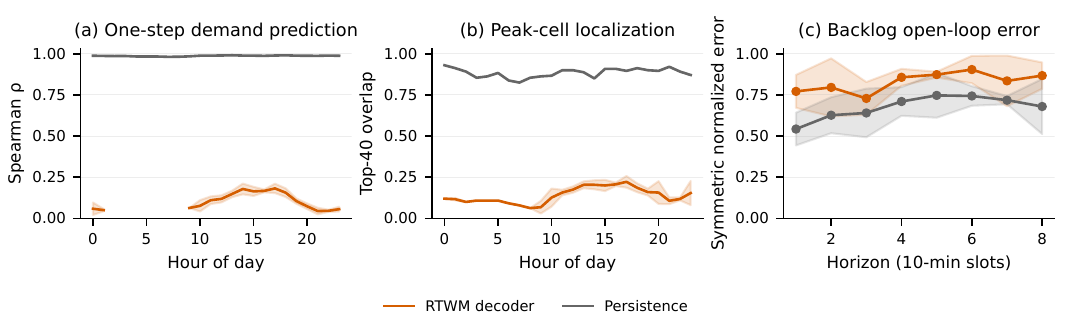}\hfill%
\includegraphics[width=0.48\columnwidth,trim=172.25bp 21bp 174.9bp 3bp,clip]{results/fig10_it1_wm.pdf}\par\smallskip
\includegraphics[width=0.48\columnwidth,trim=339.9bp 21bp 0bp 3bp,clip]{results/fig10_it1_wm.pdf}\par
\caption{Milan first-test-day open-loop diagnostics. Orange denotes RTWM and gray denotes persistence. Panels show demand rank correlation, top-40 cell overlap, and recursive queue error. Shading denotes sample SD across seeds.}
\label{fig:wm_open_loop}
\end{figure}

Fig.~\ref{fig:wm_open_loop} compares RTWM with persistence on the first Milan test day. Demand diagnostics use 143 one-step slots per run, or 429 slots in total. RTWM's Spearman correlation is defined for $199/429=46.4\%$ of slots; the remaining 53.6\% are undefined because the prediction or target is constant. Persistence is defined for $429/429$ slots. These are not common valid samples, so the Spearman averages do not support a like-for-like ranking. RTWM has lower top-40 cell overlap and higher recursive queue error through eight slots. The 11 anchors apply only to queue error, which includes inactive EUs. For predicted backlog $\widehat b$ and observed backlog $b$, the metric is $|\widehat b-b|/\max\{\widehat b,b,1~\mathrm{bit}\}$.

\ifextendedfigures
\subsection{Extended Trace and Performance Plots}
Fig.~\ref{fig:main_dsr} visualizes Table~\ref{tab:main_dsr}. Fig.~\ref{fig:perday_dsr} compares four methods on each evaluation day. Figs.~\ref{fig:secondary_metrics} and~\ref{fig:pareto} report absolute secondary metrics and empirical energy--DSR fronts. Fig.~\ref{fig:eda} shows training trace profiles and representative-EU trajectories.
\begin{figure*}[t]
\centering
\includegraphics[width=0.84\textwidth]{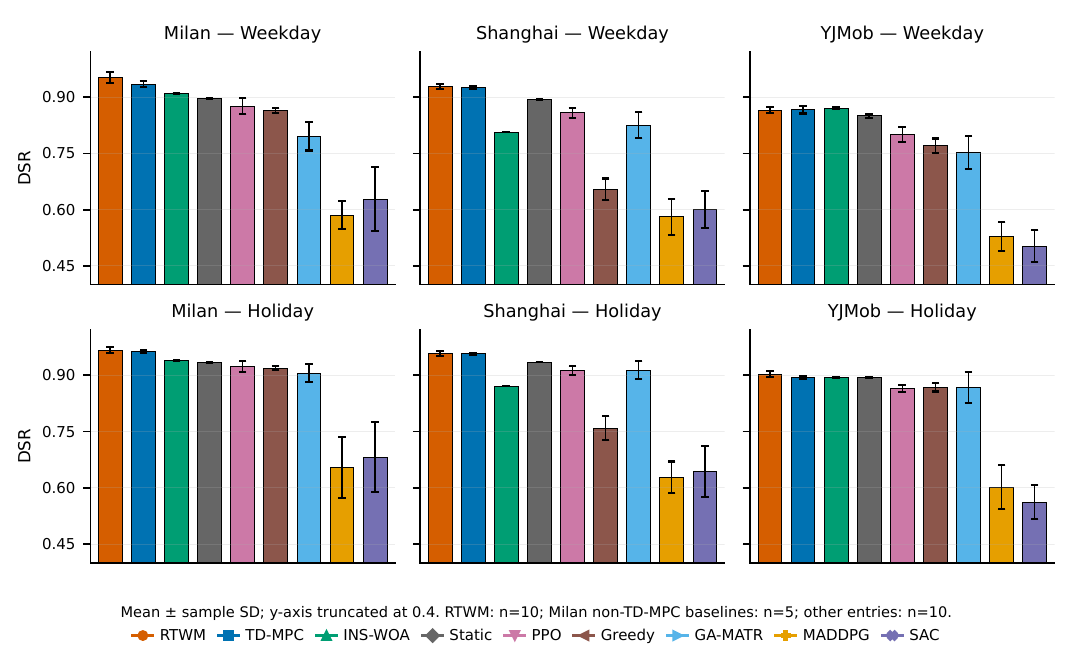}
\caption{DSR across datasets and calendar groups. Bars show means and sample SDs. The vertical scale starts at 0.4. TD-MPC denotes the compact TD-MPC-style comparator.}
\label{fig:main_dsr}
\end{figure*}
\begin{figure*}[t]
\centering
\includegraphics[width=0.84\textwidth]{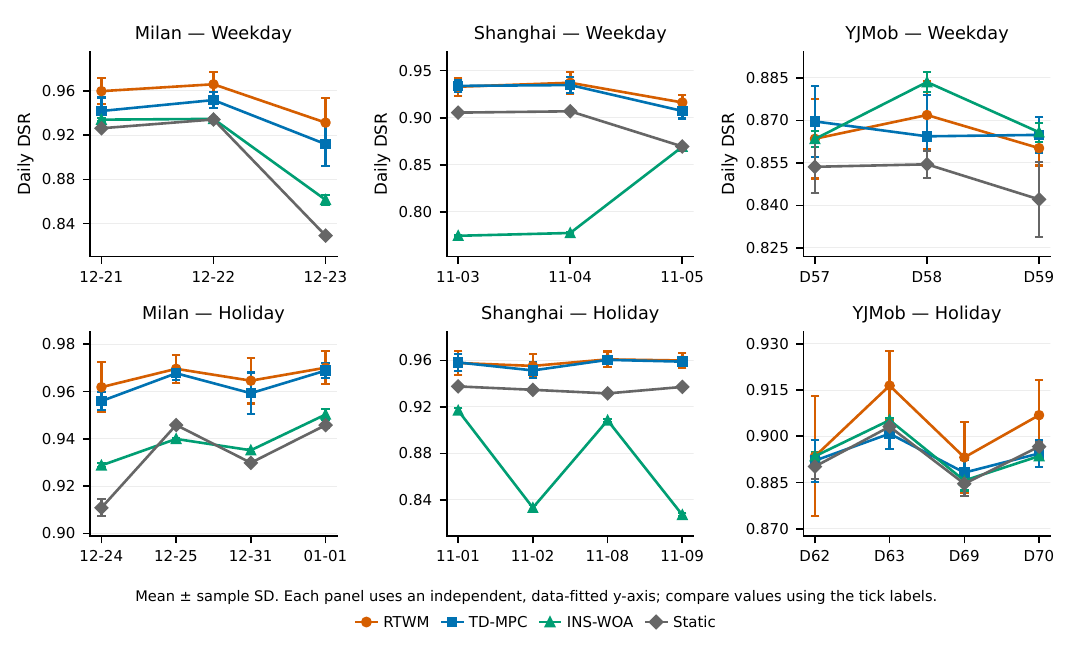}
\caption{Per-day DSR for RTWM, the TD-MPC-style comparator, INS-WOA, and Static. Markers and bars show means and sample SDs. Vertical scales vary across panels.}
\label{fig:perday_dsr}
\end{figure*}
\begin{figure*}[t]
\centering
\includegraphics[width=0.84\textwidth]{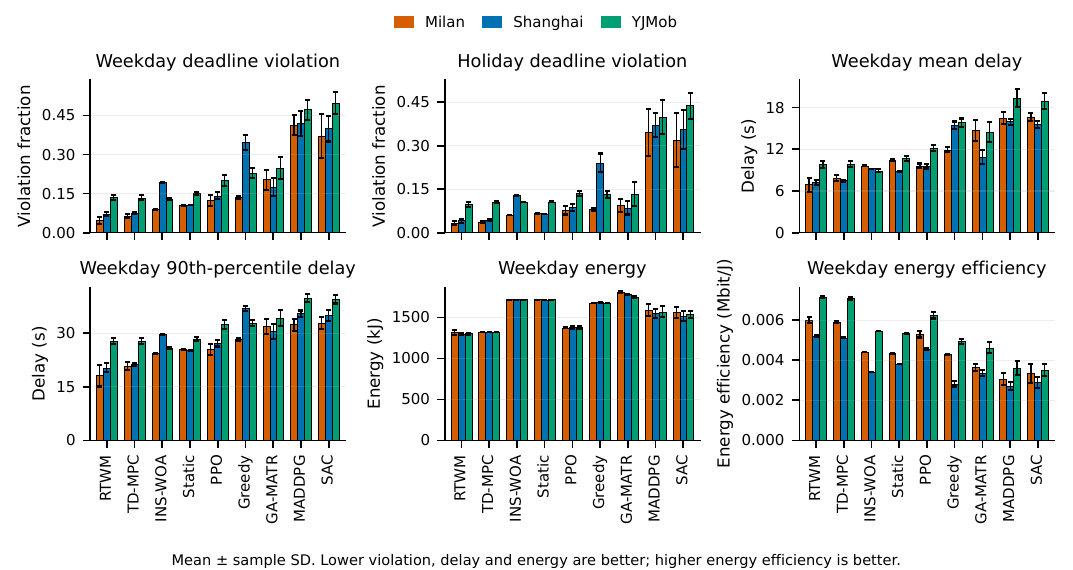}
\caption{Deadline violation by calendar group and regular-day delay, aerial-fleet energy, and energy efficiency. Bars show means and sample SDs. Delay is measured on on-time completions.}
\label{fig:secondary_metrics}
\end{figure*}
\begin{figure*}[t]
\centering
\includegraphics[width=0.84\textwidth]{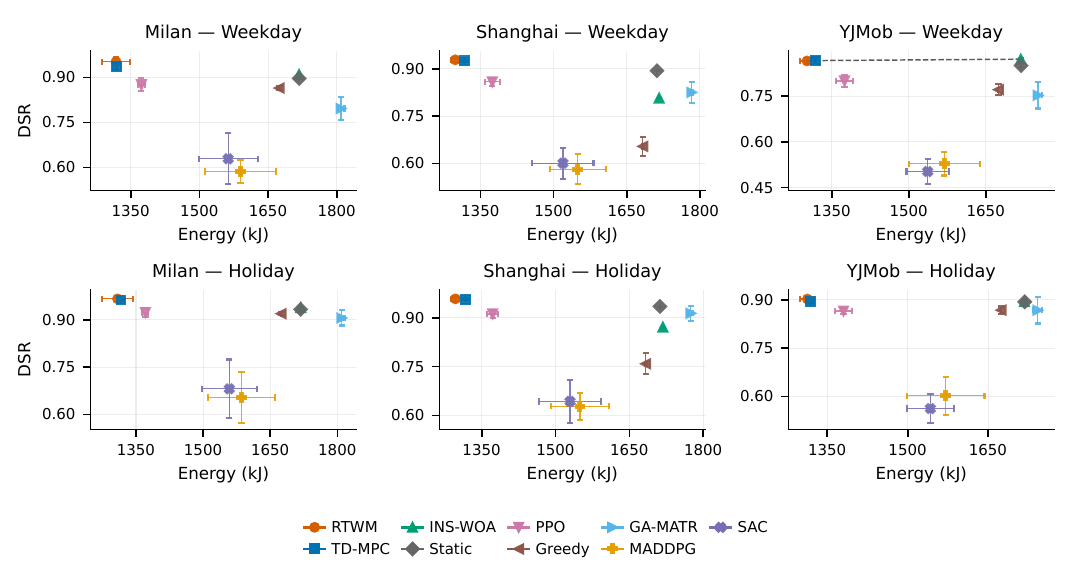}
\caption{Aerial-fleet energy versus DSR. Markers and bars show means and sample SDs. Dashed lines connect empirical Pareto points within each dataset and calendar group.}
\label{fig:pareto}
\end{figure*}
\begin{figure*}[t]
\centering
\includegraphics[width=0.82\textwidth]{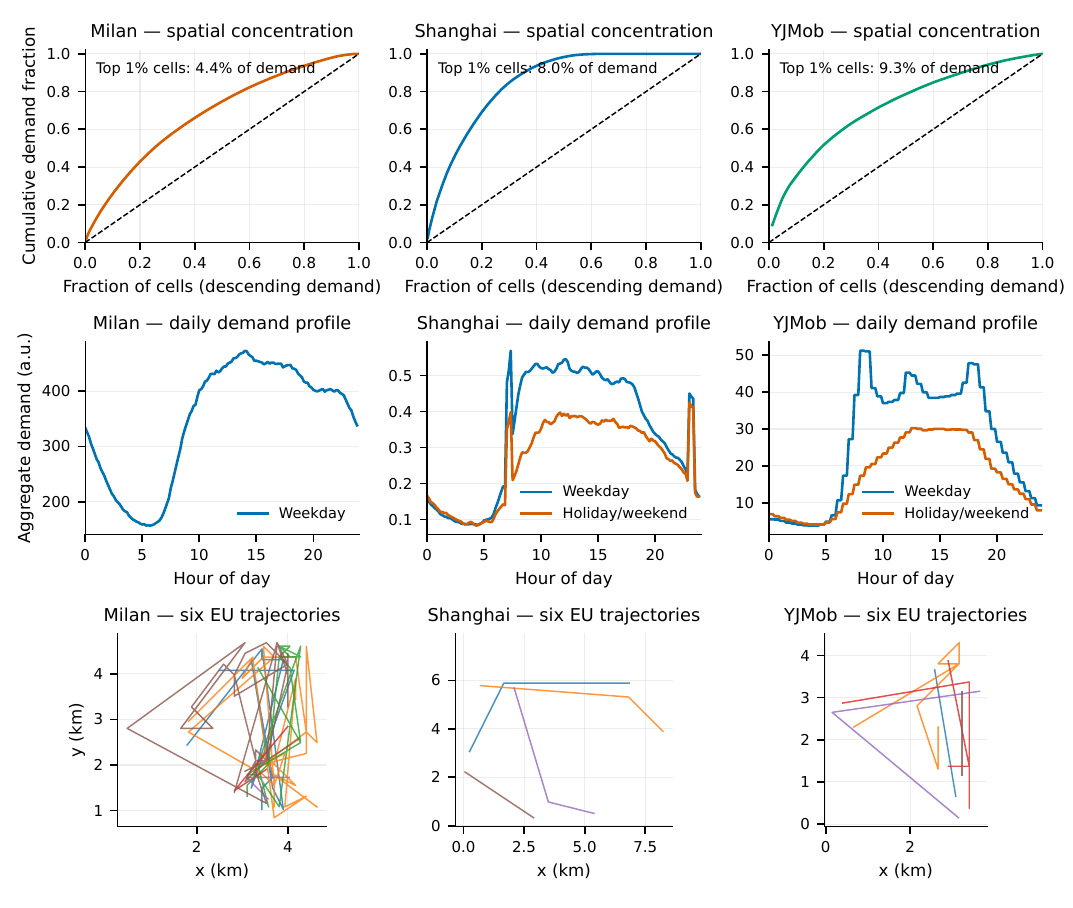}
\caption{Training trace profiles. The rows show spatial demand concentration at the peak training hour, aggregate daily demand, and six representative-EU trajectories. The dashed diagonal marks uniform spatial demand.}
\label{fig:eda}
\end{figure*}
\fi

\section{Conclusion}
\label{sec:conclusion}
RTWM shares action-free demand and EU forecasts across candidate action sequences while propagating service, queues, motion, and batteries separately. Under the evaluated budgets, its observed mean regular-day DSR is higher than the compact TD-MPC-style comparator's on Milan and Shanghai and, on YJMob, descriptively 0.1 percentage points lower---on par within seed variation. Weak open-loop forecasts remain a limit of the tested controller; at 32 UAVs, balancing the ray candidate pool with Gaussian draws restores the observed mean above both heuristics. Independent evaluation and component-matched controls would establish whether the factorization improves action selection.

\bibliographystyle{IEEEtran}
\bibliography{references}

@article{alhourani2014optimal,
  author = {Al-Hourani, Akram and Kandeepan, Sithamparanathan and Lardner, Simon},
  title = {Optimal {LAP} altitude for maximum coverage},
  journal = {IEEE Wireless Communications Letters},
  volume = {3},
  number = {6},
  pages = {569--572},
  year = {2014},
  doi = {10.1109/lwc.2014.2342736}
}

@article{zeng2019energy,
  author = {Zeng, Yong and Xu, Jie and Zhang, Rui},
  title = {Energy minimization for wireless communication with rotary-wing {UAV}},
  journal = {IEEE Transactions on Wireless Communications},
  volume = {18},
  number = {4},
  pages = {2329--2345},
  year = {2019},
  doi = {10.1109/twc.2019.2902559}
}

@article{wang2016mobileedge,
  author = {Wang, Yanting and Sheng, Min and Wang, Xijun and Wang, Liang and Li, Jiandong},
  title = {Mobile-edge computing: Partial computation offloading using dynamic voltage scaling},
  journal = {IEEE Transactions on Communications},
  volume = {64},
  number = {10},
  pages = {4268--4282},
  year = {2016},
  doi = {10.1109/tcomm.2016.2599530}
}

@article{li2022profit,
  author = {Li, Yuanzhe and Zhou, Ao and Ma, Xiao and Wang, Shangguang},
  title = {Profit-aware edge server placement},
  journal = {IEEE Internet of Things Journal},
  volume = {9},
  number = {1},
  pages = {55--67},
  year = {2022},
  doi = {10.1109/jiot.2021.3082898}
}

@article{guo2020user,
  author = {Guo, Yan and Wang, Shangguang and Zhou, Ao and Xu, Jinliang and Yuan, Jie and Hsu, Ching-Hsien},
  title = {User allocation-aware edge cloud placement in mobile edge computing},
  journal = {Software: Practice and Experience},
  volume = {50},
  number = {5},
  pages = {489--502},
  year = {2020},
  doi = {10.1002/spe.2685}
}

@article{wang2021delay,
  author = {Wang, Shangguang and Guo, Yan and Zhang, Ning and Yang, Peng and Zhou, Ao and Shen, Xuemin},
  title = {Delay-aware microservice coordination in mobile edge computing: A reinforcement learning approach},
  journal = {IEEE Transactions on Mobile Computing},
  volume = {20},
  number = {3},
  pages = {939--951},
  year = {2021},
  doi = {10.1109/tmc.2019.2957804}
}

@article{barlacchi2015multisource,
  author = {Barlacchi, Gianni and De Nadai, Marco and Larcher, Roberto and Casella, Antonio and Chitic, Cristiana and Torrisi, Giovanni and Antonelli, Fabrizio and Vespignani, Alessandro and Pentland, Alex and Lepri, Bruno},
  title = {A multi-source dataset of urban life in the city of {Milan} and the {Province of Trentino}},
  journal = {Scientific Data},
  volume = {2},
  number = {1},
  pages = {150055},
  year = {2015},
  doi = {10.1038/sdata.2015.55}
}

@misc{mexwell2023telecom,
  author = {{{mexwell}}},
  title = {Telecom {Shanghai} Dataset},
  year = {2023},
  howpublished = {Kaggle dataset mirror},
  note = {Data provided by Shanghai Telecom},
  url = {https://www.kaggle.com/datasets/mexwell/telecom-shanghai-dataset}
}

@article{yabe2024yjmob100k,
  author = {Yabe, Takahiro and Tsubouchi, Kota and Shimizu, Toru and Sekimoto, Yoshihide and Sezaki, Kaoru and Moro, Esteban and Pentland, Alex},
  title = {{YJMob100K}: City-scale and longitudinal dataset of anonymized human mobility trajectories},
  journal = {Scientific Data},
  volume = {11},
  number = {1},
  pages = {397},
  year = {2024},
  doi = {10.1038/s41597-024-03237-9}
}

@misc{schulman2017proximal,
  author = {Schulman, John and Wolski, Filip and Dhariwal, Prafulla and Radford, Alec and Klimov, Oleg},
  title = {Proximal policy optimization algorithms},
  year = {2017},
  note = {arXiv:1707.06347},
  doi = {10.48550/arXiv.1707.06347},
  url = {https://arxiv.org/abs/1707.06347}
}

@inproceedings{haarnoja2018soft,
  author = {Haarnoja, Tuomas and Zhou, Aurick and Abbeel, Pieter and Levine, Sergey},
  title = {Soft actor-critic: Off-policy maximum entropy deep reinforcement learning with a stochastic actor},
  booktitle = {Proceedings of the 35th International Conference on Machine Learning},
  series = {Proceedings of Machine Learning Research},
  volume = {80},
  pages = {1861--1870},
  year = {2018},
  publisher = {PMLR},
  url = {https://proceedings.mlr.press/v80/haarnoja18b.html}
}

@inproceedings{lowe2017multiagent,
  author = {Lowe, Ryan and Wu, Yi I. and Tamar, Aviv and Harb, Jean and Abbeel, Pieter and Mordatch, Igor},
  title = {Multi-agent actor-critic for mixed cooperative-competitive environments},
  booktitle = {Advances in Neural Information Processing Systems},
  volume = {30},
  pages = {6379--6390},
  year = {2017},
  publisher = {Curran Associates, Inc.},
  url = {https://proceedings.neurips.cc/paper/2017/hash/68a9750337a418a86fe06c1991a1d64c-Abstract.html}
}

@article{mirjalili2016whale,
  author = {Mirjalili, Seyedali and Lewis, Andrew},
  title = {The whale optimization algorithm},
  journal = {Advances in Engineering Software},
  volume = {95},
  pages = {51--67},
  year = {2016},
  doi = {10.1016/j.advengsoft.2016.01.008}
}

@inproceedings{williams2017information,
  author = {Williams, Grady and Wagener, Nolan and Goldfain, Brian and Drews, Paul and Rehg, James M. and Boots, Byron and Theodorou, Evangelos A.},
  title = {Information theoretic {MPC} for model-based reinforcement learning},
  booktitle = {2017 IEEE International Conference on Robotics and Automation (ICRA)},
  pages = {1714--1721},
  year = {2017},
  doi = {10.1109/icra.2017.7989202}
}

@misc{ha2018world,
  author = {Ha, David and Schmidhuber, J{\"u}rgen},
  title = {World models},
  year = {2018},
  note = {arXiv:1803.10122},
  doi = {10.48550/arXiv.1803.10122},
  url = {https://arxiv.org/abs/1803.10122}
}

@inproceedings{hafner2019learning,
  author = {Hafner, Danijar and Lillicrap, Timothy and Fischer, Ian and Villegas, Ruben and Ha, David and Lee, Honglak and Davidson, James},
  title = {Learning latent dynamics for planning from pixels},
  booktitle = {Proceedings of the 36th International Conference on Machine Learning},
  series = {Proceedings of Machine Learning Research},
  volume = {97},
  pages = {2555--2565},
  year = {2019},
  publisher = {PMLR},
  url = {https://proceedings.mlr.press/v97/hafner19a.html}
}

@inproceedings{hansen2022temporal,
  author = {Hansen, Nicklas A. and Su, Hao and Wang, Xiaolong},
  title = {Temporal difference learning for model predictive control},
  booktitle = {Proceedings of the 39th International Conference on Machine Learning},
  series = {Proceedings of Machine Learning Research},
  volume = {162},
  pages = {8387--8406},
  year = {2022},
  publisher = {PMLR},
  url = {https://proceedings.mlr.press/v162/hansen22a.html}
}

@inproceedings{hansen2024tdmpc2,
  author = {Hansen, Nicklas and Su, Hao and Wang, Xiaolong},
  title = {{TD-MPC2}: Scalable, robust world models for continuous control},
  booktitle = {The Twelfth International Conference on Learning Representations},
  year = {2024},
  url = {https://openreview.net/forum?id=Oxh5CstDJU}
}

@article{hafner2025mastering,
  author = {Hafner, Danijar and Pasukonis, Jurgis and Ba, Jimmy and Lillicrap, Timothy},
  title = {Mastering diverse control tasks through world models},
  journal = {Nature},
  volume = {640},
  number = {8059},
  pages = {647--653},
  year = {2025},
  doi = {10.1038/s41586-025-08744-2}
}

@article{bai2024delay,
  author = {Bai, Zhuoyi and Lin, Yifan and Cao, Yang and Wang, Wei},
  title = {Delay-aware cooperative task offloading for multi-{UAV} enabled edge-cloud computing},
  journal = {IEEE Transactions on Mobile Computing},
  volume = {23},
  number = {2},
  pages = {1034--1049},
  year = {2024},
  doi = {10.1109/tmc.2022.3232375}
}

@article{hao2024joint,
  author = {Hao, Hao and Xu, Changqiao and Zhang, Wei and Yang, Shujie and Muntean, Gabriel-Miro},
  title = {Joint task offloading, resource allocation, and trajectory design for multi-{UAV} cooperative edge computing with task priority},
  journal = {IEEE Transactions on Mobile Computing},
  volume = {23},
  number = {9},
  pages = {8649--8663},
  year = {2024},
  doi = {10.1109/tmc.2024.3350078}
}

@article{feng2024graphattention,
  author = {Feng, Zikai and Wu, Di and Huang, Mengxing and Yuen, Chau},
  title = {Graph-attention-based reinforcement learning for trajectory design and resource assignment in multi-{UAV}-assisted communication},
  journal = {IEEE Internet of Things Journal},
  volume = {11},
  number = {16},
  pages = {27421--27434},
  year = {2024},
  doi = {10.1109/jiot.2024.3397823}
}

@article{jia2026dynamic,
  author = {Jia, Ziye and He, Jia and He, Lijun and Sheng, Min and Liu, Junyu and Wu, Qihui and Han, Zhu},
  title = {Dynamic trajectory optimization and power control for hierarchical {UAV} swarms in {6G} aerial access network},
  journal = {IEEE Transactions on Wireless Communications},
  volume = {25},
  pages = {3349--3362},
  year = {2026},
  doi = {10.1109/twc.2025.3603432}
}

@article{cao2024computational,
  author = {Cao, Pan and Lei, Lei and Cai, Shengsuo and Shen, Gaoqing and Liu, Xiaojiao and Wang, Xinyi and Zhang, Lijuan and Zhou, Liang and Guizani, Mohsen},
  title = {Computational intelligence algorithms for {UAV} swarm networking and collaboration: A comprehensive survey and future directions},
  journal = {IEEE Communications Surveys \& Tutorials},
  volume = {26},
  number = {4},
  pages = {2684--2728},
  year = {2024},
  doi = {10.1109/comst.2024.3395358}
}

@article{sun2024multiobjective,
  author = {Sun, Geng and Wang, Yixian and Sun, Zemin and Wu, Qingqing and Kang, Jiawen and Niyato, Dusit and Leung, Victor C. M.},
  title = {Multi-objective optimization for multi-{UAV}-assisted mobile edge computing},
  journal = {IEEE Transactions on Mobile Computing},
  volume = {23},
  number = {12},
  pages = {14803--14820},
  year = {2024},
  doi = {10.1109/tmc.2024.3446819}
}

@article{zhang2024task,
  author = {Zhang, Yuhao and Kuang, Zhufang and Feng, Yanyan and Hou, Fen},
  title = {Task offloading and trajectory optimization for secure communications in dynamic user multi-{UAV} {MEC} systems},
  journal = {IEEE Transactions on Mobile Computing},
  volume = {23},
  number = {12},
  pages = {14427--14440},
  year = {2024},
  doi = {10.1109/tmc.2024.3442909}
}

@article{peng2025trajectory,
  author = {Peng, Sicong and Li, Bin and Liu, Lei and Fei, Zesong and Niyato, Dusit},
  title = {Trajectory design and resource allocation for multi-{UAV}-assisted sensing, communication, and edge computing integration},
  journal = {IEEE Transactions on Communications},
  volume = {73},
  number = {4},
  pages = {2847--2861},
  year = {2025},
  doi = {10.1109/tcomm.2024.3478115}
}

@article{sun2025tjcct,
  author = {Sun, Zemin and Sun, Geng and Wu, Qingqing and He, Long and Liang, Shuang and Pan, Hongyang and Niyato, Dusit and Yuen, Chau and Leung, Victor C. M.},
  title = {{TJCCT}: A two-timescale approach for {UAV}-assisted mobile edge computing},
  journal = {IEEE Transactions on Mobile Computing},
  volume = {24},
  number = {4},
  pages = {3130--3147},
  year = {2025},
  doi = {10.1109/tmc.2024.3505155}
}

@article{pervez2024energy,
  author = {Pervez, Farhan and Sultana, Ajmery and Yang, Cungang and Zhao, Lian},
  title = {Energy and latency efficient joint communication and computation optimization in a multi-{UAV}-assisted {MEC} network},
  journal = {IEEE Transactions on Wireless Communications},
  volume = {23},
  number = {3},
  pages = {1728--1741},
  year = {2024},
  doi = {10.1109/twc.2023.3291692}
}

@article{guo2024multiuav,
  author = {Guo, Hongzhi and Wang, Yutao and Liu, Jiajia and Liu, Chang},
  title = {Multi-{UAV} cooperative task offloading and resource allocation in {5G} advanced and beyond},
  journal = {IEEE Transactions on Wireless Communications},
  volume = {23},
  number = {1},
  pages = {347--359},
  year = {2024},
  doi = {10.1109/twc.2023.3277801}
}

@article{ning2025joint,
  author = {Ning, Zhaolong and Ji, Hongjing and Wang, Xiaojie and Ngai, Edith C. H. and Guo, Lei and Liu, Jiangchuan},
  title = {Joint optimization of data acquisition and trajectory planning for {UAV}-assisted wireless powered {Internet of Things}},
  journal = {IEEE Transactions on Mobile Computing},
  volume = {24},
  number = {2},
  pages = {1016--1030},
  year = {2025},
  doi = {10.1109/tmc.2024.3470831}
}

@article{li2026joint,
  author = {Li, Yihang and Gao, Xiaozheng and Zhang, Zeyu and Yuan, Hang and Kang, Jiawen and Niyato, Dusit and Yang, Kai},
  title = {Joint trajectory, resource, and access optimization in multi-{UAV} collaborative mobile edge computing networks for low-altitude economy},
  journal = {IEEE Internet of Things Journal},
  volume = {13},
  number = {5},
  pages = {9467--9481},
  year = {2026},
  doi = {10.1109/jiot.2025.3645730}
}

@article{wu2025fairness,
  author = {Wu, Jintao and Xu, Xiaolong and Cui, Guangming and Zhang, Yiwen and Qi, Lianyong and Dou, Wanchun and Cai, Zhipeng},
  title = {Fairness-aware budgeted edge server placement for connected autonomous vehicles},
  journal = {IEEE Transactions on Mobile Computing},
  volume = {24},
  number = {6},
  pages = {4762--4776},
  year = {2025},
  doi = {10.1109/tmc.2025.3526873}
}

@article{zhang2024dvfo,
  author = {Zhang, Ziyang and Zhao, Yang and Li, Huan and Lin, Changyao and Liu, Jie},
  title = {{DVFO}: Learning-based {DVFS} for energy-efficient edge-cloud collaborative inference},
  journal = {IEEE Transactions on Mobile Computing},
  volume = {23},
  number = {10},
  pages = {9042--9059},
  year = {2024},
  doi = {10.1109/tmc.2024.3357218}
}

@article{zhao2025joint,
  author = {Zhao, Mingxiong and Zhang, Rongqian and He, Zhenli and Li, Keqin},
  title = {Joint optimization of trajectory, offloading, caching, and migration for {UAV}-assisted {MEC}},
  journal = {IEEE Transactions on Mobile Computing},
  volume = {24},
  number = {3},
  pages = {1981--1998},
  year = {2025},
  doi = {10.1109/tmc.2024.3486995}
}

@article{chen2025dynamic,
  author = {Chen, Yang and Pi, Dechang and Yang, Shengxiang and Xu, Yue and Wang, Bi and Qin, Shuo and Wang, Yintong},
  title = {A dynamic optimization framework for computation rate maximization in {UAV}-assisted mobile edge computing},
  journal = {IEEE Transactions on Vehicular Technology},
  volume = {74},
  number = {7},
  pages = {11395--11409},
  year = {2025},
  doi = {10.1109/tvt.2025.3546026}
}

@article{hao2025reliability,
  author = {Hao, Hao and Xu, Changqiao and Zhang, Wei and Chen, Xingyan and Yang, Shujie and Muntean, Gabriel-Miro},
  title = {Reliability-aware optimization of task offloading for {UAV}-assisted edge computing},
  journal = {IEEE Transactions on Computers},
  volume = {74},
  number = {11},
  pages = {3832--3844},
  year = {2025},
  doi = {10.1109/tc.2025.3604463}
}

@article{du2024enhancing,
  author = {Du, Hongyang and Zhang, Ruichen and Liu, Yinqiu and Wang, Jiacheng and Lin, Yijing and Li, Zonghang and Niyato, Dusit and Kang, Jiawen and Xiong, Zehui and Cui, Shuguang and Ai, Bo and Zhou, Haibo and Kim, Dong In},
  title = {Enhancing deep reinforcement learning: A tutorial on generative diffusion models in network optimization},
  journal = {IEEE Communications Surveys \& Tutorials},
  volume = {26},
  number = {4},
  pages = {2611--2646},
  year = {2024},
  doi = {10.1109/comst.2024.3400011}
}

@article{wang2025joint,
  author = {Wang, Yuhui and Farooq, Junaid and Ghazzai, Hakim and Setti, Gianluca},
  title = {Joint positioning and computation offloading in multi-{UAV} {MEC} for low latency applications: A proximal policy optimization approach},
  journal = {IEEE Transactions on Mobile Computing},
  volume = {24},
  number = {10},
  pages = {9584--9598},
  year = {2025},
  doi = {10.1109/tmc.2025.3562806}
}

@article{deng2024entropy,
  author = {Deng, Tan and Wang, Yanping and Li, Jin and Cao, Ronghui and Gu, Yongtong and Hu, Jinming and Tang, Xiaoyong and Huang, Mingfeng and Liu, Wenzheng and Li, Shixue},
  title = {Entropy normalization {SAC}-based task offloading for {UAV}-assisted mobile-edge computing},
  journal = {IEEE Internet of Things Journal},
  volume = {11},
  number = {15},
  pages = {26220--26233},
  year = {2024},
  doi = {10.1109/jiot.2024.3395276}
}

@article{kim2024cooperative,
  author = {Kim, Mintae and Lee, Hoon and Hwang, Sangwon and Debbah, M{\'e}rouane and Lee, Inkyu},
  title = {Cooperative multiagent deep reinforcement learning methods for {UAV}-aided mobile edge computing networks},
  journal = {IEEE Internet of Things Journal},
  volume = {11},
  number = {23},
  pages = {38040--38053},
  year = {2024},
  doi = {10.1109/jiot.2024.3447090}
}

@article{hwang2025multiagent,
  author = {Hwang, Sangwon and Lee, Hoon and Kim, Mintae and Lee, Inkyu},
  title = {Multiagent deep reinforcement learning for decentralized multi-{UAV} mobile edge computing networks},
  journal = {IEEE Internet of Things Journal},
  volume = {12},
  number = {10},
  pages = {14484--14497},
  year = {2025},
  doi = {10.1109/jiot.2025.3527016}
}

@article{tarekegn2025trajectory,
  author = {Tarekegn, Getaneh Berie and Tesfaw, Belayneh Abebe and Juang, Rong-Terng and Saha, Dola and Tarekegn, Robel Berie and Lin, Hsin-Piao and Tai, Li-Chia},
  title = {Trajectory control and fair communications for multi-{UAV} networks: A federated multi-agent deep reinforcement learning approach},
  journal = {IEEE Transactions on Wireless Communications},
  volume = {24},
  number = {9},
  pages = {7598--7611},
  year = {2025},
  doi = {10.1109/twc.2025.3561271}
}

@article{xing2026multiuav,
  author = {Xing, Xiaojun and Ma, Yuanqiang and Lei, Yichen and Li, Yan and Xiao, Bing},
  title = {Multi-{UAV} rendezvous trajectory planning based on improved {MADDPG} algorithm in complex dynamic obstacle environments},
  journal = {IEEE Transactions on Vehicular Technology},
  volume = {75},
  number = {4},
  pages = {5580--5591},
  year = {2026},
  doi = {10.1109/tvt.2025.3624052}
}

@article{tang2025deep,
  author = {Tang, Xiao and Zhao, Kexin and Shen, Chao and Du, Qinghe and Wang, Yichen and Niyato, Dusit and Han, Zhu},
  title = {Deep graph reinforcement learning for {UAV}-enabled multi-user secure communications},
  journal = {IEEE Transactions on Mobile Computing},
  volume = {24},
  number = {9},
  pages = {8780--8793},
  year = {2025},
  doi = {10.1109/tmc.2025.3558790}
}

@article{liu2024graph,
  author = {Liu, Ziheng and Zhang, Jiayi and Shi, Enyu and Liu, Zhilong and Niyato, Dusit and Ai, Bo and Shen, Xuemin},
  title = {Graph neural network meets multi-agent reinforcement learning: Fundamentals, applications, and future directions},
  journal = {IEEE Wireless Communications},
  volume = {31},
  number = {6},
  pages = {39--47},
  year = {2024},
  doi = {10.1109/mwc.015.2300595}
}

@article{gong2025sttf,
  author = {Gong, Jiahui and Liu, Yu and Li, Tong and Ding, Jingtao and Wang, Zhaocheng and Jin, Depeng},
  title = {{STTF}: A spatiotemporal transformer framework for multi-task mobile network prediction},
  journal = {IEEE Transactions on Mobile Computing},
  volume = {24},
  number = {5},
  pages = {4072--4085},
  year = {2025},
  doi = {10.1109/tmc.2024.3521245}
}

@inproceedings{he2025mamba,
  author = {He, Sicheng and Ji, Junzhong and Lei, Minglong},
  title = {Decomposed spatio-temporal {Mamba} for long-term traffic prediction},
  booktitle = {Proceedings of the AAAI Conference on Artificial Intelligence},
  volume = {39},
  number = {11},
  pages = {11772--11780},
  year = {2025},
  doi = {10.1609/aaai.v39i11.33281}
}

@article{zhou2025cooperative,
  author = {Zhou, Longyu and Leng, Supeng and Wang, Qing and Quek, Tony Q.S. and Guizani, Mohsen},
  title = {Cooperative digital twins for {UAV}-based scenarios},
  journal = {IEEE Communications Magazine},
  volume = {63},
  number = {3},
  pages = {40--46},
  year = {2025},
  doi = {10.1109/mcom.001.2400207}
}

@article{shen2024large,
  author = {Shen, Yifei and Shao, Jiawei and Zhang, Xinjie and Lin, Zehong and Pan, Hao and Li, Dongsheng and Zhang, Jun and Letaief, Khaled B.},
  title = {Large language models empowered autonomous edge {AI} for connected intelligence},
  journal = {IEEE Communications Magazine},
  volume = {62},
  number = {10},
  pages = {140--146},
  year = {2024},
  doi = {10.1109/mcom.001.2300550}
}

@article{guan2024world,
  author = {Guan, Yanchen and Liao, Haicheng and Li, Zhenning and Hu, Jia and Yuan, Runze and Li, Yunjian and Zhang, Guohui and Xu, Chengzhong},
  title = {World models for autonomous driving: An initial survey},
  journal = {IEEE Transactions on Intelligent Vehicles},
  pages = {1--17},
  year = {2024},
  doi = {10.1109/tiv.2024.3398357},
  note = {Early access}
}

@inproceedings{frauenknecht2024trust,
  author = {Frauenknecht, Bernd and Eisele, Artur and Subhasish, Devdutt and Solowjow, Friedrich and Trimpe, Sebastian},
  title = {Trust the model where it trusts itself---model-based actor-critic with uncertainty-aware rollout adaption},
  booktitle = {Proceedings of the 41st International Conference on Machine Learning},
  series = {Proceedings of Machine Learning Research},
  volume = {235},
  pages = {13973--14005},
  year = {2024},
  publisher = {PMLR},
  url = {https://proceedings.mlr.press/v235/frauenknecht24a.html}
}

@inproceedings{wang2025world,
  author = {Wang, Lingyi and Shelim, Rashed and Saad, Walid and Ramakrishnan, Naren},
  title = {World model-based learning for long-term age of information minimization in vehicular networks},
  booktitle = {2025 IEEE Global Communications Conference (GLOBECOM)},
  pages = {6202--6207},
  year = {2025},
  doi = {10.1109/globecom59602.2025.11431797}
}

@article{andrejev2025pimppi,
  author = {Andrejev, Edvin Martin and Manoharan, Amith and Unt, Karl-Eerik and Singh, Arun Kumar},
  title = {{$\pi$-MPPI}: A projection-based model predictive path integral scheme for smooth optimal control of fixed-wing aerial vehicles},
  journal = {IEEE Robotics and Automation Letters},
  volume = {10},
  number = {6},
  pages = {6496--6503},
  year = {2025},
  doi = {10.1109/lra.2025.3568323}
}
\end{document}